\documentclass[aps,prc,showpacs]{revtex4}

\usepackage{epsfig}

\newcommand{\vsig}{\mbox{\boldmath$\sigma$\unboldmath}}

\newcommand{\be}{\begin{equation}}
\newcommand{\ee}{\end{equation}}
\newcommand{\bea}{\begin{eqnarray}}
\newcommand{\eea}{\end{eqnarray}}
\newcommand{\bean}{\begin{eqnarray*}}
\newcommand{\eean}{\end{eqnarray*}}

\newcommand{\gapproxeq}{\lower
.7ex\hbox{$\;\stackrel{\textstyle >}{\sim}\;$}}
\newcommand{\lapproxeq}{\lower
.7ex\hbox{$\;\stackrel{\textstyle <}{\sim}\;$}}

\begin{document}

\title{Charmed baryon strong decays in a chiral quark model }
\author{
Xian-Hui Zhong$^{1,3}$
and
Qiang Zhao$^{1,2,3}$
}

\affiliation{ 1) Institute of High Energy Physics,
       Chinese Academy of Sciences, Beijing 100049, P.R. China
}
\affiliation{ 2) Department of Physics, University of Surrey,
Guildford, GU2 7XH, United Kingdom
 }
\affiliation{ 3) Theoretical Physics Center for Science Facilities
(TPCSF), Chinese Academy of Sciences, Beijing 100049, P.R. China
}


\begin{abstract}

Charmed baryon strong decays are studied in a chiral quark model.
The data for the decays of $\Lambda^+_c(2593)$,
$\Lambda^+_c(2625)$, $\Sigma^{++,+,0}_c$ and
$\Sigma^{+,0}_c(2520)$, are accounted for successfully, which
allows to fix the pseudoscalar-meson-quark couplings in an
effective chiral Lagrangian. Extending this framework to analyze
the strong decays of the newly observed charmed baryons, we
classify that both $\Lambda_c(2880)$ and $\Lambda_c(2940)$ are
$D$-wave states in the $N=2$ shell; $\Lambda_c(2880)$ could be
$|\Lambda_c \ ^2 D_{\lambda\lambda}\frac{3}{2}^+\rangle $ and
$\Lambda_c(2940)$ could be $|\Lambda_c \ ^2
D_{\lambda\lambda}\frac{5}{2}^+\rangle $. Our calculation also
suggests that $\Lambda_c(2765)$ is very likely a $\rho$-mode
$P$-wave excited state in the $N=1$ shell, and favors a
$|\Lambda_c \ ^4 P_\rho\frac{1}{2}^-\rangle$ configuration. The
$\Sigma_c(2800)$ favors being a $|\Sigma_c \ ^2
P_\lambda\frac{1}{2}^-\rangle $ state. But its being
$|\Sigma^{++}_c \ ^4 P_\lambda\frac{5}{2}^-\rangle $ cannot be
ruled out.
\end{abstract}


\maketitle

\section{Introduction}

In the past years, some new charmed baryons, such as
$\Lambda_c(2880)$, $\Lambda_c(2940)$, $\Lambda_c(2765)$ and
$\Sigma^{++,+,0}_c(2800)$,  were observed in Belle, BaBar and CLEO
\cite{Aubert:2006sp,Abe:2006rz,Mizuk:2004yu,Artuso:2000xy}. It
initiated great interests in the heavy flavor baryon spectrum in
both experiment and theory. At present, the experimental
information is still limited and nearly nothing is known for their
spin-parity quantum numbers (some review of the charmed baryons
can be found in
\cite{Cheng:2006dk,Chen:2007xf,Kreps:2007vy,Rosner:2006jz}). How
to understand the properties of these new charmed baryons, e.g.
their structures, and interactions with known particles in their
production and decay, and how to establish the charmed baryon
spectroscopy, have been hot topics in both experiment and theory.

By studying the transitions of their different decay modes, one
expects to extract information about their structures and the
underlying dynamics. Several classes of models have been developed
to deal with the strong decays of baryons. One is the hadrodynamic
model, in which all hadrons are treated as point-like objects. The
heavy hadron chiral perturbation theory (HHChPT) approach
\cite{Cheng:2006dk,Huang:1995ke} belongs to this class. The second
class of models treats the baryons as a three quark system, while
the meson behaves as a point-like particle emitted from active
quarks when the initial baryon decays. Typical models of this
class were review in Refs. \cite{Koniuk:1979vy,Copley:1979wj}. The
third class of models is the pair creation model, in which both
the baryon and meson have internal structures. The decay of a
hadron is recognized by the creation of a quark-antiquark pair
from vacuum, which combine with the initial quarks to form meson
and baryon in the final state. Typical ways of treating the pair
creations include the $^3P_0$ model
\cite{LeYaouanc:1974mr,Chen:2007xf}, string-breaking model
\cite{Dosch:1986dp,Alcock:1983gb}, and flux-tube breaking models
\cite{Isgur:1984bm,Kokoski:1985is,Kumano:1988ga,Stancu:1988gb}.
Detailed review of these phenomenologies can be found in Ref.
\cite{Capstick:1992th}. A large number of recent papers have been
contributed to the determination of the quantum numbers of these
newly observed
states~\cite{Cheng:2006dk,Chen:2007xf,Migura:2006ep,Ebert:2007nw,
Gerasyuta:2007un,Albertus:2005zy,Garcilazo:2007eh,Hwang:2006df,Ivanov:1999bk,
Tawfiq:1998nk,Huang:1995ke,Zhu:2000py,Cho:1994vg}.

In this work, we will analyze the strong decays of charmed baryons
in the non-relativistic chiral quark model, which belongs to the
second class of models, and had been well developed and widely
used for the processes of meson photo-productions
\cite{qk1,qk2,qkk,Li:1997gda,qkk2,qk3,qk4,qk5}. An extension of
this approach to describe the process of $\pi N$ scattering also
turns out to be successful and inspiring \cite{Zhong:2007fx}. In
this framework, the charmed baryon spatial wave functions are
described by harmonic oscillators. An effective chiral Lagrangian
is then introduced to account for the quark-meson coupling. Since
the quark-meson coupling is invariant under the chiral
transformation, some of the low-energy properties of QCD are
retained \cite{Zhong:2007fx,Li:1997gda,qk3}. This approach is
similar to that used in \cite{Koniuk:1979vy,Copley:1979wj}, the
only difference is that two constants in the decay amplitudes in
Refs. \cite{Koniuk:1979vy,Copley:1979wj} are replaced by two
energy-dependent factors deduced from the chiral Lagrangian in our
model.

In this work, we will first study the strong decays of the well
determined charmed baryons, $\Lambda^+_c(2593)$,
$\Lambda^+_c(2625)$, $\Sigma^{++,+,0}_c $ and
$\Sigma^{++,+,0}_c(2520)$. Using the measurement of
$\Sigma^{++}_c(2520)$ as an input, we then determine the only free
parameter $\delta$ in our model, with which we calculate the
strong decays of $\Lambda^+_c(2593)$, $\Lambda^+_c(2625)$,
$\Sigma^{++,+,0}_c $ and $\Sigma^{+,0}_c(2520)$ as a prediction.
By comparing with the data we can extract information about these
states, in particular, about these structures and quantum
numbers~\cite{PDG}.

Finally, we analyze the strong decays of the new observed charmed
baryons $\Lambda_c(2880)$, $\Lambda_c(2940)$, $\Lambda_c(2765)$
and $\Sigma^{++,+,0}_c(2800)$. We predict that both
$\Lambda_c(2880)$ and $\Lambda_c(2940)$ are $D$-wave states in
$N=2$ shell. $\Lambda_c(2880)$ could be a $|\Lambda_c \ ^2
D_{\lambda\lambda}\frac{3}{2}^+\rangle $ state and
$\Lambda_c(2940)$ could be a $|\Lambda_c \ ^2
D_{\lambda\lambda}\frac{5}{2}^+\rangle$ state. We suggest that
$\Lambda_c(2765)$ is most likely a $\rho$-mode $P$-wave
excitations charmed baryon in $N=1$ shell. The most possible state
is $|\Lambda_c \ ^4 P_\rho\frac{1}{2}^-\rangle$. The calculations
indicate that $\Sigma_c(2800)$ favors a $|\Sigma_c \ ^2 P_\lambda
\frac{1}{2}^-\rangle$ state over the other ones for its broad
width in experiment.

The paper is organized as follows. In the subsequent section, the
charmed baryons in the quark model is outlined. Then, the non-relativistic
quark-meson couplings are given in Sec.\ \ref{qmc}. The decay amplitudes
are deduced in Sec.\ \ref{apt}. We present our
calculations and discussions in Sec.\ \ref{cd}. Finally, a summary
is given in Sec.\ \ref{sum}.

\section{Charmed baryons in the quark model }

\subsection{oscillator states}
For a $udc$ basis state, it contains two light quarks 1 and 2 with
equal mass $m$, and a heavy charmed quark 3 with mass $m'$. The
basis states are generated by the Hamiltonian \cite{quarkm}
\begin{equation} \label{hm1}
\mathcal{H}=\frac{1}{2m}(\mathbf{p}^2_1+\mathbf{p}^2_2)+\frac{1}{2m'}\mathbf{p}^2_3+
\frac{1}{2}K\sum_{i< j}(\mathbf{r}_i-\mathbf{r}_j)^2.
\end{equation}
In the above non-relativistic expansions, vectors $\textbf{r}_{j}$
and $\textbf{p}_j$ are the coordinate and momentum for the $j$-th
quark in the baryon rest frame. The quarks are confined in an
oscillator potential with the potential parameter $K$ independent
of the flavor quantum number. One defines the Jacobi coordinates
to eliminate the c.m. variables:
\begin{eqnarray}
\vec{\rho}&=&\frac{1}{\sqrt{2}}(\mathbf{r}_1-\mathbf{r}_2),\label{zb1}\\
\vec{\lambda}&=&\frac{1}{\sqrt{6}}(\mathbf{r}_1+\mathbf{r}_2-2\mathbf{r}_3),\label{zb2}\\
\mathbf{R}_{c.m.}&=&\frac{m(\mathbf{r}_1+\mathbf{r}_2)+m'\mathbf{r}_3}{2m+m'}\label{zb3}.
\end{eqnarray}
With the above relations (\ref{zb1}--\ref{zb3}), the oscillator
Hamiltonian (\ref{hm1}) is reduced to
\begin{eqnarray} \label{hm2}
\mathcal{H}=\frac{P^2_{cm}}{2 M}+\frac{1}{2m_\rho}\mathbf{p}^2_\rho+\frac{1}{2m_\lambda}\mathbf{p}^2_\lambda+
\frac{3}{2}K(\rho^2+\lambda^2).
\end{eqnarray}
where
\begin{eqnarray} \label{mom}
\mathbf{p}_\rho=m_\rho\dot{\vec{\rho}},\ \
\mathbf{p}_\lambda=m_\lambda\dot{\vec{\lambda}},\ \
\mathbf{P}_{c.m.}=M \mathbf{\dot{R}}_{c.m.},
\end{eqnarray}
with
\begin{eqnarray} \label{mass}
M=2m+m',\ \ m_\rho=m,\ \ m_\lambda=\frac{3m m'}{2m+m'}.
\end{eqnarray}
With Eqs.(\ref{zb1}--\ref{zb3}) and (\ref{mom}) the coordinate
$\mathbf{r}_j$ can be translated into functions of the Jacobi
coordinates $\lambda$ and $\rho$:
\begin{eqnarray}
\mathbf{r}_1&=&\mathbf{R}_{c.m.}+\frac{1}{\sqrt{6}}\frac{3m'}{2m+m'}\vec{\lambda}+\frac{1}{\sqrt{2}}\vec{\rho},\\
\mathbf{r}_2&=&\mathbf{R}_{c.m.}+\frac{1}{\sqrt{6}}\frac{3m'}{2m+m'}\vec{\lambda}-\frac{1}{\sqrt{2}}\vec{\rho},\\
\mathbf{r}_3&=&\mathbf{R}_{c.m.}-\sqrt{\frac{2}{3}}\frac{3m}{2m+m'}\vec{\lambda},
\end{eqnarray}
and the momentum $\mathbf{p}_j$ is given by
\begin{eqnarray}
\mathbf{p}_1&=&\frac{m}{M}\mathbf{P}_{c.m.}+\frac{1}{\sqrt{6}}\mathbf{p}_\lambda+\frac{1}{\sqrt{2}}\mathbf{p}_\rho,\\
\mathbf{p}_2&=&\frac{m}{M}\mathbf{P}_{c.m.}+\frac{1}{\sqrt{6}}\mathbf{p}_\lambda-\frac{1}{\sqrt{2}}\mathbf{p}_\rho,\\
\mathbf{p}_3&=&\frac{m'}{M}\mathbf{P}_{c.m.}-\sqrt{\frac{2}{3}}\mathbf{p}_\lambda.
\end{eqnarray}

The spatial wave function is a product of the $\rho$-oscillator
and the $\lambda$-oscillator states. With the standard notation,
the principal quantum numbers of the $\rho$-oscillator and
$\lambda$-oscillator are $N_\rho=(2n_\rho+l_\rho)$ and
$N_\lambda=(2n_\lambda+l_\lambda)$, and  the energy of a state is
given by
\begin{eqnarray}
E_N&=&(N_\rho+\frac{3}{2})\omega_\rho+(N_\lambda+\frac{3}{2})\omega_\lambda
\ .
\end{eqnarray}
The total principal quantum number (i.e. shell number) $N$ is
defined as
\be
N=N_\rho+N_\lambda,
\ee
and the frequencies of the $\rho$-mode and $\lambda$-mode are
\begin{eqnarray}\label{freq}
\omega_\rho=(3K/m_\rho)^{1/2},\ \
\omega_\lambda=(3K/m_\lambda)^{1/2}.
\end{eqnarray}
In the quark model two useful oscillator parameters, i.e. the
potential strengths, are defined by
\begin{eqnarray} \label{par}
\alpha_\rho=(m_\rho \omega_\rho)^{1/2}, \ \
\alpha_\lambda=(m_\lambda \omega_\lambda)^{1/2}.
\end{eqnarray}
Combining Eqs.(\ref{mass}) and (\ref{freq}) with (\ref{par}), we
obtain the relation between these two parameters:
\begin{eqnarray} 
\alpha^2_\lambda=\sqrt{\frac{3m'}{2m+m'}}\alpha^2_\rho.
\end{eqnarray}

Then, the wave function of an oscillator is give by
\begin{eqnarray}
\psi^{n_\sigma}_{l_\sigma m}(\sigma)=R_{n_\sigma
l_\sigma}(\sigma)Y_{l_\sigma m}(\sigma),
\end{eqnarray}
where $\sigma=\rho,\lambda$.  The total orbital angular momentum
$\mathbf{L}$ of a state is obtained by coupling $\mathbf{l}_\rho$
to $\mathbf{l}_\lambda$:
\begin{eqnarray}
\mathbf{L}=\mathbf{l}_\rho+\mathbf{l}_\lambda.
\end{eqnarray}
The total spatial wave function can then be constructed. All the
functions with principal quantum number $N\leq 2$ are listed in
Tab. \ref{wffff}.

\subsection{Flavor and spin wave functions}

For the $udc$ basis states which violate $SU(4)$ symmetry, as done
in Ref. \cite{Copley:1979wj} we introduce
\begin{eqnarray}
\phi_{\Lambda_c}=\frac{1}{\sqrt{2}}(ud-du)c,
\end{eqnarray}
and
\begin{eqnarray}
\phi_{\Sigma_c}=\cases{d d c & for $\Sigma^{0}_c$\cr
\frac{1}{\sqrt{2}}(ud+du)c  & for $\Sigma^{+}_c$\cr u u c & for
$\Sigma^{++}_c$},
\end{eqnarray}
for the $\Lambda_c$- and $\Sigma_c$-type flavor wave functions,
respectively.

For the spin wave functions  the usual ones are adopted
\cite{Koniuk:1979vy,Copley:1979wj}:
\begin{eqnarray}
\chi^s_{3/2}&=&\uparrow\uparrow\uparrow, \ \  \chi^s_{-3/2}=\downarrow\downarrow\downarrow, \nonumber\\
\chi^s_{1/2}&=&\frac{1}{\sqrt{3}}(\uparrow\uparrow\downarrow+\uparrow\downarrow\uparrow+\downarrow\uparrow\uparrow),\nonumber\\
\chi^s_{-1/2}&=&\frac{1}{\sqrt{3}}(\uparrow\downarrow\downarrow+\downarrow\downarrow\uparrow+\downarrow\uparrow\downarrow),
\end{eqnarray}
for the spin-3/2 states;
\begin{eqnarray}
\chi^\rho_{1/2}&=&\frac{1}{\sqrt{2}}(\uparrow\downarrow\uparrow-\downarrow\uparrow\uparrow),\nonumber\\
\chi^\rho_{-1/2}&=&\frac{1}{\sqrt{2}}(\uparrow\downarrow\downarrow-\downarrow\uparrow\downarrow),
\end{eqnarray}
for the spin-1/2 states, in which the first two quark spins are
antisymmetric; and
\begin{eqnarray}
\chi^\lambda_{1/2}&=&-\frac{1}{\sqrt{6}}(\uparrow\downarrow\uparrow+\downarrow\uparrow\uparrow-2\uparrow\uparrow\downarrow),\nonumber\\
\chi^\lambda_{-1/2}&=&-\frac{1}{\sqrt{6}}(\uparrow\downarrow\downarrow+\downarrow\uparrow\downarrow-2\downarrow\downarrow\uparrow),
\end{eqnarray}
for the spin-1/2 states, in which the first two quark spins are
symmetric.

\subsection{The total wave functions}

The spin-flavor and spatial wave functions of baryons must be
symmetric since the color wave function is antisymmetric. The
flavor wave functions of the $\Lambda_c$-type charmed baryons,
$\phi_{\Lambda_c}$, are antisymmetric under the interchange of the
$u$ and $d$ quarks, thus, their spin-space wave functions must be
symmetric. In contrast, the spin-spatial wave functions of
$\Sigma_c$-type charmed baryons are required to be antisymmetric
due to their symmetric flavor wave functions under the interchange
of the $u$ and $d$ quarks. The wave functions of the
$\Lambda_c$-type and $\Sigma_c$-type charmed baryons are listed in
Tabs. \ref{wfL} and \ref{wfS} respectively.

\section{The quark-meson couplings }\label{qmc}

In the chiral quark model, the low energy quark-meson interactions
are described by an effective Lagrangian \cite{Li:1997gda,qk3}
\be \label{lg}
\mathcal{L}=\bar{\psi}[\gamma_{\mu}(i\partial^{\mu}+V^{\mu}+\gamma_5A^{\mu})-m]\psi
+\cdot\cdot\cdot,
\ee
where $V^{\mu}$ and $A^{\mu}$ correspond to
vector and axial currents, respectively. They are given by
\begin{eqnarray}
V^{\mu} &=&
 \frac{1}{2}(\xi\partial^{\mu}\xi^{\dag}+\xi^{\dag}\partial^{\mu}\xi),
\nonumber\\
 A^{\mu}
&=&
 \frac{1}{2i}(\xi\partial^{\mu}\xi^{\dag}-\xi^{\dag}\partial^{\mu}\xi),
\end{eqnarray}
with $ \xi=\exp{(i \phi_m/f_m)}$, where $f_m$ is the meson decay
constant. In the flavor $SU(3)$ sector, the pseudoscalar-meson
octet $\phi_m$ can be expressed as
\begin{eqnarray}
\phi_m=\pmatrix{
 \frac{1}{\sqrt{2}}\pi^0+\frac{1}{\sqrt{6}}\eta & \pi^+ & K^+ \cr
 \pi^- & -\frac{1}{\sqrt{2}}\pi^0+\frac{1}{\sqrt{6}}\eta & K^0 \cr
 K^- & \bar{K}^0 & -\sqrt{\frac{2}{3}}\eta},
\end{eqnarray}
and the quark field $\psi$ is given by
\begin{eqnarray}\label{qf}
\psi=\pmatrix{\psi(u)\cr \psi(d) \cr \psi(s) }.
\end{eqnarray}

The tree-level quark-meson pseudovector coupling is thus given by
\begin{eqnarray}\label{coup}
H_m=\sum_j
\frac{1}{f_m}\bar{\psi}_j\gamma^{j}_{\mu}\gamma^{j}_{5}\psi_j\partial^{\mu}\phi_m.
\end{eqnarray}
where $\psi_j$ represents the $j$-th quark field in a baryon. This
effective quark-meson pseudovector coupling can be used for
$D$-mesons as well, if we extend the $SU(3)$ case to the $SU(4)$
case.

In the quark model, the non-relativistic form of Eq. (\ref{coup})
is written as \cite{Zhong:2007fx,Li:1997gda,qk3}
\begin{eqnarray}\label{non-relativistic-expans}
H^{nr}_{m}&=&\sum_j\Big\{\frac{\omega_m}{E_f+M_f}\vsig_j\cdot
\textbf{P}_f+ \frac{\omega_m}{E_i+M_i}\vsig_j \cdot
\textbf{P}_i \nonumber\\
&&-\vsig_j \cdot \textbf{q} +\frac{\omega_m}{2\mu_q}\vsig_j\cdot
\textbf{p}'_j\Big\}I_j \varphi_m,
\end{eqnarray}
where $\vsig_j$ and $\mu_q$ correspond to the Pauli spin vector
and the reduced mass of the $j$-th quark in the initial and final
baryons, respectively. For emitting a meson, we have
$\varphi_m=e^{-i\textbf{q}\cdot \textbf{r}_j}$, and for absorbing
a meson we have $\varphi_m=e^{i\textbf{q}\cdot \textbf{r}_j}$. In
the above non-relativistic expansions,
$\textbf{p}'_j=\textbf{p}_j-(m_j/M) \textbf{P}_{c.m.}$ is the
internal momentum for the $j$-th quark in the baryon rest frame.
$\omega_m$ and $\textbf{q}$ are the energy and three-vector
momentum of the meson, respectively. The isospin operator $I_j$ in
Eq. (\ref{non-relativistic-expans}) is expressed as
\begin{eqnarray}
I_j=\cases{a^{\dagger}_j(u)a_j(c) & for $D^0$\cr
a^{\dagger}_j(u)a_j(d) & for $\pi^-$\cr a^{\dagger}_j(d)a_j(u)  &
for $\pi^+$\cr
\frac{1}{\sqrt{2}}[a^{\dagger}_j(u)a_j(u)-a^{\dagger}_j(d)a_j(d)]
& for $\pi^0$},
\end{eqnarray}
where $a^{\dagger}_j(u,d,c)$ and $a_j(u,d,c)$ are the creation and
annihilation operators for the $u$, $d$ and $c$ quarks.

\section{The decay of charmed baryon in the quark model}\label{apt}

In the calculations, we select the initial-baryon-rest system for
the decay precesses. The energies and momenta of the initial charmed
baryons are denoted by $(E_i, \textbf{P}_i$), while those of the
final state mesons and baryons are denoted by $(\omega_f,
\textbf{q})$ and $(E_f, \textbf{P}_f)$. Note that $\textbf{P}_i=0$
($E_i=M_i$) and $\textbf{P}_f=-\textbf{q}$.

\subsection{$\mathcal{B}_c\to  \mathcal{B}'_c \pi(\mathbf{q})$}

Because the $\pi$-meson only couples to the light quark 1 or 2 in a
$udc$ basis state, the strong decay amplitudes for the process
$\mathcal{B}_c\to  \mathcal{B}'_c \pi(\mathbf{q})$ can be written as
\begin{eqnarray} \label{am}
&& \mathcal{M}[\mathcal{B}_c\to   \mathcal{B}'_c
\pi(\mathbf{q})]  \nonumber\\
& =& 2\left\langle \mathcal{B}'_c\left|\left\{G\vsig_1\cdot
 \textbf{q}
+h\vsig_1\cdot \textbf{p}'_1\right\}I_1 e^{-i\textbf{q}\cdot
\textbf{r}_1}\right|\mathcal{B}_c\right\rangle ,
\end{eqnarray}
with
\begin{eqnarray}\label{ccpk}
G\equiv -\frac{\omega_\pi}{E_f+M_f}-1,\ \ h\equiv
\frac{\omega_\pi}{m},
\end{eqnarray}
where $\mathcal{B}_c$ and $\mathcal{B}'_c$ stand for the initial and
final charmed baryon wave functions, which are listed in Tabs.
\ref{wfL} and \ref{wfS}. Similar expressions were also derived in
Refs. \cite{Koniuk:1979vy,Copley:1979wj}. By selecting
$\mathbf{q}=q\hat{z}$, namely the meson moves along the $z$ axial,
we can simplify the amplitude to
\begin{eqnarray} \label{samm}
\mathcal{M}[\mathcal{B}_c &\to  &  \mathcal{B}'_c \pi(\mathbf{q})]\nonumber\\
&=&2\left\{Gq-\frac{1}{\sqrt{2}}\left(\frac{1}{\sqrt{3}}q_\lambda+q_\rho\right)h\right\}\langle
\mathcal{B}'_c|\sigma_{1z}\phi I_1|\mathcal{B}_c \rangle\nonumber\\
&&-i\sqrt{\frac{2}{3}}h\langle \mathcal{B}'_c|(\vsig_1\cdot
\vec{\nabla}_\lambda-\alpha^2_\lambda\vsig_1\cdot\vec{\lambda})\phi
I_1|
\mathcal{B}_c\rangle\nonumber\\
&&-i\sqrt{2}h\langle \mathcal{B}'_c|(\vsig_1\cdot
\vec{\nabla}_\rho-\alpha^2_\rho\vsig_1\cdot\vec{\rho})\phi I_1|
\mathcal{B}_c\rangle,
\end{eqnarray}
where $\vec{\nabla}_\lambda$ and $\vec{\nabla}_\rho$ are the
derivative operators on the spatial wave function of the final
baryon except the factor
$\mathrm{exp}[(-\alpha^2_\lambda\lambda^2-\alpha^2_\rho\rho^2)/2]$
which has been worked out, and
\begin{eqnarray}
q_\lambda=\frac{1}{\sqrt{6}}\frac{3m'}{2m+m'}q,\ \
q_\rho=\frac{1}{\sqrt{2}}q,
\end{eqnarray}
and
\begin{eqnarray}
\phi=\exp(-iq_\lambda \lambda_z) \exp(-iq_\rho \rho_z),
\end{eqnarray}
In Eq. (\ref{samm}), the first term comes from the c.m. motion of
the system, while the last two terms attribute to the $\lambda$-
and $\rho$-mode orbital excitations of the charmed baryons,
respectively.

For example, we calculate the decay process $|\Lambda_c \ ^2
P_\lambda\frac{1}{2}^-\rangle\to  \Sigma_c \pi$. The initial and
final charmed baryon wave functions are given by (see Tab.
\ref{wfL})
\begin{eqnarray}
|\mathcal{B}_c\rangle&=&\left[\sqrt{\frac{1}{3}}\Psi^{\lambda}_{11}
\chi^{\rho}_{-1/2} +\sqrt{\frac{2}{3}}\Psi^{\lambda}_{10}
\chi^{\rho}_{1/2} \right]\phi_{\Lambda_c},\\
|\mathcal{B}'_c\rangle&=& \Psi^S_{00}\chi^\lambda_{1/2}\phi_{\Sigma_c}.
\end{eqnarray}
Substituting into Eq.(\ref{samm}), we obtain the decay amplitude
\begin{eqnarray}
\mathcal{M}&=& ig_1g_I  \bigg\{ \sqrt{\frac{2}{3}}\left[
Gq-\frac{h}{2\sqrt{2}}\left(\frac{1} {\sqrt{3}}q_\lambda+q_\rho
\right) \right]\frac{q_\lambda}
{\alpha_\lambda}  \nonumber\\
&&  + h\alpha_\lambda \bigg \}F(q_\lambda,q_\rho) ,
\end{eqnarray}
where the spin and isospin factors are
\begin{eqnarray}
g_1=\langle \chi^\lambda_{1/2}|\sigma_{1z}|\chi^\rho_{1/2} \rangle,
\end{eqnarray}
and
\begin{eqnarray}
g_I=\langle \phi_{\Sigma_c}|\sigma_{1z}|\phi_{\Lambda_c} \rangle.
\end{eqnarray}
The spatial integral gives
\begin{equation} \label{formf}
F(q_\lambda,q_\rho)=\exp
\left(-\frac{q^2_\lambda}{4\alpha^2_\lambda}-\frac{q^2_\rho}{4\alpha^2_\rho}\right),
\end{equation}
which plays a role of form factor.

The corresponding spin factors are listed in Tab. \ref{gfactor}.
Some of the decay amplitudes for $|\Lambda_c \ ^{2S+1}
L_{\sigma}J^P\rangle\to \Sigma_c \pi$, $|\Lambda_c \ ^{2S+1}
L_{\sigma}J^P\rangle\to \Sigma_c(2520) \pi$, $|\Sigma_c \ ^{2S+1}
L_{\sigma}J^P\rangle\to  \Lambda_c \pi$, $|\Sigma_c \ ^{2S+1}
L_{\sigma}J^P\rangle\to  \Sigma_c \pi$ and $|\Sigma_c \ ^{2S+1}
L_{\sigma}J^P\rangle\to  \Sigma_c(2520) \pi$ are listed in Tabs.
\ref{asa},  \ref{asb}, \ref{css}, \ref{F12} and \ref{f11},
respectively.

\subsection{$\mathcal{B}_c\to    D(\mathbf{q})p $}\label{t}

For a charmed baryon decaying into $D p $, since the $D$-meson only
couples to the charm quark 3 in a $udc$ basis state, the strong
decay amplitudes for the process $\mathcal{B}_c\to D(\mathbf{q})p$
can be written as
\begin{eqnarray} \label{amd}
&& \mathcal{M}[\mathcal{B}_c  \to  D(\mathbf{q})p]\nonumber\\
&=&\left\langle
p\left|\left\{G\vsig_3\cdot
 \textbf{q}
+h\vsig_3\cdot \textbf{p}'_3\right\}I_3 e^{-i\textbf{q}\cdot
\textbf{r}_3}\right|\mathcal{B}_c\right\rangle, \nonumber\\
\end{eqnarray}
where the wave function of a proton in the quark model is expressed as
\begin{eqnarray}
|p\rangle =\frac{1}{\sqrt{2}}(\Phi_\rho
\chi^\rho+\Phi_\lambda\chi^\lambda) \ ^0\Psi^S_{00},
\end{eqnarray}
with
\begin{eqnarray}
\Phi_\rho &=&\frac{1}{\sqrt{2}}(ud-du)u, \\
\Phi_\lambda&=&-\frac{1}{\sqrt{6}}(udu+duu-2uud).
\end{eqnarray}
We can also simplify the amplitude (\ref{amd}) to
\begin{eqnarray} \label{samd}
\mathcal{M}[\mathcal{B}_c &\to  & D(\mathbf{q})p]
=\left[Gq-\sqrt{\frac{2}{3}}q'_\lambda h\right]\langle
p|\sigma_{3z}\phi' I_3|\mathcal{B}_c \rangle\nonumber\\
&&+i\sqrt{\frac{2}{3}}  h\langle p|(\vsig_3\cdot
\vec{\nabla}_\lambda-\alpha^2_\lambda\vsig_3\cdot\vec{\lambda})\phi'
I_3| \mathcal{B}_c\rangle.
\end{eqnarray}
with
\begin{eqnarray}\label{ccpk-1}
G\equiv -\frac{\omega_D}{E_f+M_f}-1,\ \ h\equiv \omega_D
\frac{m+m'}{2m'm},
\end{eqnarray}
and
\begin{eqnarray}
q'_\lambda=\frac{2}{\sqrt{6}}\frac{3m}{2m+m'}q,\ \
\phi'=\exp(iq'_\lambda \lambda_z).
\end{eqnarray}
In Eq. (\ref{samd}), the first term comes from the c.m. motion
of the system, the last term attributes to the $\lambda$-mode
orbital excitations of the charmed baryons, respectively.

There exist selection rules for the $D^0 p$ decay channel of
$\Lambda_c$ excitations, in which only $|\Lambda_c \ ^2
D_{\lambda\lambda}\frac{3}{2}^+\rangle$, $|\Lambda_c \ ^2
D_{\lambda\lambda}\frac{5}{2}^+\rangle$ and $|\Lambda_c \ ^2
S_{\lambda\lambda}\frac{1}{2}^+\rangle$ can decay into $D^0p$.
Their decay amplitudes are listed in Tab. \ref{adp}. States of
$|\Lambda_c \ ^2 P_\lambda\frac{1}{2}^-\rangle$ and $|\Lambda_c \
^2 P_\lambda\frac{3}{2}^-\rangle$ are likely below the $D^0 p$
threshold, while others are forbidden by the spin-isospin
selection rule.

\section{calculation and analysis}\label{cd}

With the resonance decay amplitudes, one can calculate the width
\begin{equation}\label{dww}
\Gamma=\left(\frac{\delta}{f_m}\right)^2\frac{(E_f+M_f)|\textbf{q}|}{4\pi
M_i }\frac{1}{2J_i+1}
\sum_{J_{iz},J_{fz}}|\mathcal{M}_{J_{iz},J_{fz}}|^2 ,
\end{equation}
where $J_i$ and $J_f$ are the total  angular momenta of the
initial and final baryons, respectively. A dimensionless constant,
$\delta$, is introduced to take into account uncertainties arising
from the model and to be determined by experimental data. In the
calculation, the standard parameters of the quark model are
adopted. For the oscillator parameters, we use
$\alpha^2_\rho=0.16$ GeV$^2$. The $u$, $d$ constituent quark
masses are $m=350$ MeV, and the charm quark mass is $m'=1700$ MeV.
The decay constants for $\pi$- and $D$-mesons are $f_\pi=132$ MeV
and $f_D=226$ MeV, which are taken from the Particle Data Group
(PDG) \cite{PDG}. All the charmed baryon masses are also adopted
from the PDG \cite{PDG}.

\subsection{$\Sigma_c $ and $\Sigma_c(2520)$}

$\Sigma_c $ and $\Sigma_c(2520)$ are the two lowest states in the
$\Sigma_c$-type charmed baryons. They are assigned to the two
$S$-wave states, $|\Sigma_c \ ^2 S\frac{1}{2}^+\rangle$ and
$|\Sigma_c \ ^4 S\frac{3}{2}^+\rangle$, respectively \cite{PDG,
Copley:1979wj}. We use the measured width for
$\Sigma^{++}_c(2520)\to \Lambda^+_c \pi^+$ as an input ( i.e.
$\Gamma=14.9$ MeV) to determine parameter $\delta$ in
Eq.(\ref{dww}), which gives
\begin{eqnarray}
\delta=0.557.
\end{eqnarray}

Applying this value for $\delta$, we can predict the other strong
decay widths. In particular, the decay widths of $\Sigma_c \to
\Lambda_c \pi$, $\Sigma^{+}_c(2520)\to \Lambda^+_c \pi^0$ and
$\Sigma^{0}_c(2520)\to  \Lambda^+_c \pi^-$ are calculated. The
results are listed in Tab. \ref{wssss}, from which we find that
our predictions are in a good agreement with the experimental data
\cite{PDG}, and compatible with other theoretical predictions
\cite{Cheng:2006dk,Chen:2007xf,Ivanov:1999bk,Tawfiq:1998nk,Huang:1995ke,Albertus:2005zy}.
We also see that the decay width of $\Sigma_c(2520)$ is larger
than that of $\Sigma_c $ by a factor of $\sim 7$ though their
decay amplitudes have the same form (see Tab. \ref{css}). The
reasons are due to: i) the spin factor $g^{\Sigma}_3$ for
$\Sigma_c(2520)$ is larger than $g^{\Sigma}_1$ for $\Sigma_c$ by a
factor $\sqrt{2}$; ii) the three-momentum of the pion, $|{\bf q}|$
in the $\Sigma_c(2520) \to  \Lambda_c \pi$ are about two times
larger than that in the $\Sigma_c \to  \Lambda_c \pi$. It leads to
larger values for quantities $G$ and $h$. This feature was also
mentioned in Ref. \cite{Cheng:2006dk}.

\subsection{ $\Lambda_c(2593)$ and $\Lambda_c(2625)$}

$\Lambda_c(2593)$ and $\Lambda_c(2625)$ have $J^P=1/2^-$ and
$J^P=3/2^-$, respectively, and can be naturally assigned to $N=1$
shell with one unit of orbital angular momentum excitation. They
can be excited via either $P_\lambda$-mode or $P_\rho$-mode. For
the former assignment, their spatial wave functions are
$|\Lambda_c \ ^2 P_\lambda\frac{1}{2}^-\rangle$ and $|\Lambda_c \
^2 P_\lambda\frac{3}{2}^-\rangle$, from which the decay widths can
be calculated. As shown in Tab. \ref{wssss}, the results are in a
good agreement with the experimental data \cite{PDG} and
consistent with the classification of Ref. \cite{Copley:1979wj} in
the quark model.

Assuming $\Lambda_c(2593)$ and $\Lambda_c(2625)$ are $P_\rho$-mode
excitations, we also calculate their widths. In contrast with the
$P_\rho$-mode, they turn out to be much broader than the
$P_\lambda$-mode. For $\Lambda_c(2593)$ it is possible that the
physical state is a mixture of the $P_\lambda$- and $P_\rho$-modes
within the uncertainties of the present data though the
determination of the mixing angle will also rely on the mass of
the second state.

For $\Lambda_c(2625)$ the $P_\rho$-mode excitation turns to
overestimate the data significantly. The experimental upper limit
is about two orders of magnitude smaller than the predictions from
the $P_\rho$-mode excitation, while the $P_\lambda$-mode results
are consistent with the data. This could be a sign that the mixing
between the $P_\lambda$- and $P_\rho$-mode in $\Lambda_c(2625)$
should be small. Concerning the possible mixings between the
$P_\lambda$- and $P_\rho$-mode excitations, the search for the
second heavier $1/2^-$ and $3/2^-$ states in experiment should be
interesting.

Comparing $\Lambda_c(2593)$ with $\Lambda_c(2625)$, it shows that
the decay width of $\Lambda_c(2593)$ is much narrower than that of
$\Lambda_c(2625)$, which can be well understood in our model. In
the decay amplitude of $\Lambda_c(2625)\to  \Sigma_c \pi$ (see
Tab. \ref{asa}), only c.m. motion contributions are present, which
leads to the small decay width. We should also emphasize that the
partial decay width of $\Lambda_c(2593)\to \Sigma_c\pi$ is
sensitive to the mass of $\pi$-meson due to its mass close to the
$\Sigma_c\pi$ threshold. It leads to the decay width of $\Sigma_c
\pi^0$ channel is about two times larger than those of
$\Sigma_c\pi^{\pm}$. Interestingly, experimental data for
$\Lambda_c(2593)$ and $\Lambda_c(2625)\to \Sigma_c^+ \pi^0$ are
still not available.

Since the well-determined $S$- and $P$-wave charmed baryon strong
decay widths are successfully described in our chiral quark model,
we extend this approach in the next subsections to investigate the
strong decays of other newly observed charmed baryons, such as
$\Lambda_c(2880)$ and $\Lambda_c(2940)$.

\subsection{$\Lambda_c(2880)$}

$\Lambda_c(2880)$ was observed in $\Lambda^+_c \pi^+\pi^-$ by CLEO
\cite{Artuso:2000xy}, in $D^0p$ channel by BaBar
\cite{Aubert:2006sp}, and in $\Sigma_c\pi$, $\Sigma_c(2520)\pi$ by
Belle \cite{Abe:2006rz}. It has a narrow decay width less than 8
MeV \cite{PDG,Abe:2006rz}, based on which it was proposed to be a
$\tilde{\Lambda}^+_{c0}(\frac{1}{2}^-)$ state in Ref.
\cite{Artuso:2000xy}. In the heavy hadron chiral perturbation
theory, Cheng {\it et al.} made a conjecture that
$\Lambda_c(2880)$ is an admixture of $\Lambda_{c2}(\frac{5}{2}^+)$
with $\tilde{\Lambda}^{''}_{c3}(\frac{5}{2}^+)$
\cite{Cheng:2006dk} which are both $L=2$ orbitally excited states.
Chen {\it et al.} suggested that $\Lambda_c(2880)$ favors
$\tilde{\Lambda}^{2}_{c3}(\frac{5}{2}^+)$ within the $^3P_0$ model
\cite{Chen:2007xf}. According to the quark model predictions, the
mass for $J^P=3/2^+$ is around 2.9 GeV, which indicates
$\Lambda_c(2880)$ maybe favor $J^P=3/2^+$ as well
\cite{Capstick:1986bm,Ebert:2007nw}. The other suggestions about
its quantum numbers also can be found in Ref.
\cite{Garcilazo:2007eh}.

Meanwhile, the Belle measurement~\cite{Abe:2006rz} shows
contributions from intermediate $\Sigma_c^* $ states in
$\Lambda^+_c(2880)\to \Sigma^*_c \pi\to \Lambda^+_c\pi^+\pi^-$,
and the ratio of the partial decay widths for the intermediate
$\Sigma_c(2520)$ and $\Sigma_c $ is extracted:
\begin{eqnarray}
\mathcal{R}=\frac{\Gamma(\Sigma_c(2520)\pi)}{\Gamma(\Sigma_c
\pi)}= 0.225\pm 0.062\pm 0.025.
\end{eqnarray}
With the analysis of the angular distributions in
$\Lambda_c(2880)\to \Sigma^{0,++}_c  \pi^{+,-}$ decays, the
$\Lambda_c(2880)$ spin-parity assignment is favored to be
$J^P=5/2^+$ over the others.

In the quark model the masses of $N=1$ shell $\Lambda_c$
excitations are at the order of $2.5$-2.6 GeV, which is much less
than 2.88 GeV. We hence only consider the possible assignment of
$\Lambda_c(2880)$ in the $N=2$ shell. As shown by Tab.
\ref{w2880}, only $|\Lambda_c \ ^2
D_{\lambda\lambda}\frac{3}{2}^+\rangle $ can produce results that
fit in the three experimental observations: i) with a narrow decay
width; ii) decaying into $D^0p$; iii) and with the ratio
$\mathcal{R}=\Gamma(\Sigma_c(2520)\pi)/\Gamma(\Sigma_c \pi)\simeq
0.25$. This is an  orbital excitation with $l_\lambda=2$ and
$l_\rho=0$. Note that the Capstick-Isgur quark model
\cite{Capstick:1986bm} and the relativistic quark model
\cite{Ebert:2007nw} predict the lowest $J^P=3/2^+$ $\Lambda_c$
excitation at 2910 MeV and $2874$ MeV, respectively, which are
consistent with the experimental value within the model
accuracies. In this sense the assignment of $\Lambda_c(2880)$ as
$|\Lambda_c \ ^2 D_{\lambda\lambda}\frac{3}{2}^+\rangle$ turns to
be possible.

Interestingly, the experimental analysis of the decay angular
distribution \cite{Abe:2006rz} indicates a preference of
$J^P=5/2^+$ over $3/2^+$ at a level of more than 4.5 standard
deviations. By assigning $5/2^+$ to the $\Lambda_c(2880)$, we find
that only the state $|\Lambda_c \ ^2 D_A\frac{5}{2}^+\rangle$ is
close to the experimental measurements. However, its $D^0p$ decay
channel is forbidden and the ratio $\mathcal{R}=0.5$ turns to be
too large compared with the Belle data. This controversy may
suggest that $\Lambda_c(2880)$ is neither a pure $|\Lambda_c \ ^2
D_{\lambda\lambda}\frac{3}{2}^+\rangle$ nor $|\Lambda_c \ ^2
D_A\frac{5}{2}^+\rangle$. We expect that more accurate
measurements of the decay angular distributions will clarify its
nature. In contrast, the calculations of Refs.
\cite{Cheng:2006dk,Chen:2007xf} seem to agree with the data. The
details of our model calculations are listed in Tab.~\ref{w2880}.

\subsection{$\Lambda_c(2940)$}

$\Lambda_c(2940)$ was first seen in its decay into $D^0p$ by BaBar
\cite{Aubert:2006sp}, and then confirmed by Belle in
$\Sigma_c^{0,++} \pi^{+,-}$ \cite{Abe:2006rz}. Its spin-parity has
not yet been determined. In this mass region, it can be
$J^P=5/2^+$, $J^P=3/2^+$, $J^P=1/2^+$ or $J^P=5/2^-$ as suggested
by the quark model \cite{Capstick:1986bm}. The $^3P_0$ model
\cite{Chen:2007xf} suggests that its configuration favors
$\breve{\Lambda}^{0}_{c1}(\frac{1}{2}^+)$ or
$\breve{\Lambda}^{0}_{c1}(\frac{3}{2}^+)$, while a molecular state
with $J^P=1/2^-$ is also proposed~\cite{He:2006is}.

In our analysis it shows that only three states, $|\Lambda_c \ ^2
D_{\lambda\lambda}\frac{3}{2}^+\rangle $, $|\Lambda_c \ ^2
D_{\lambda\lambda}\frac{5}{2}^+\rangle $ and $|\Lambda_c \ ^2
S_{\lambda\lambda}\frac{1}{2}^+\rangle $, can decay into $D^0p$.
In case that we have assigned $\Lambda_c(2880)$ to be $|\Lambda_c
\ ^2 D_{\lambda\lambda}\frac{3}{2}^+\rangle $, the
$\Lambda_c(2940)$ could thus be either $|\Lambda_c \ ^2
D_{\lambda\lambda}\frac{5}{2}^+\rangle $ or $|\Lambda_c \ ^2
S_{\lambda\lambda}\frac{1}{2}^+\rangle $. In the quark model, the
mass of $|\Lambda_c \ ^2 S_{\lambda\lambda}\frac{1}{2}^+\rangle $
should be less than that of $|\Lambda_c \ ^2
D_{\lambda\lambda}\frac{3}{2}^+\rangle $ [i.e. $\Lambda_c(2880)$].
This leaves $\Lambda_c(2940)$ to be assigned as $|\Lambda_c \ ^2
D_{\lambda\lambda}\frac{5}{2}^+\rangle $.

In Tab. \ref{w2940}, the calculation results are listed. The
vanishing $D^0 p$ channel will eliminate most of those states,
especially, with anti-symmetric spatial wavefunctions and mixed
$\rho\rho$-type. The states which have nonvanishing decays into
$\Sigma_c \pi$, $\Sigma_c(2520)\pi$ and $D^0 p$ are $|\Lambda_c \
^2 D_{\lambda\lambda}\frac{5}{2}^+\rangle $ and $|\Lambda_c \ ^2
S_{\lambda\lambda}\frac{1}{2}^+\rangle $. Based on the argument
made in the last paragraph, we see that it is natural to assign
the $\Lambda_c(2940)$ as $|\Lambda_c \ ^2
D_{\lambda\lambda}\frac{5}{2}^+\rangle $. Note that the
Capstick-Isgur quark model predicts the lowest $J^P=5/2^+$ state
at 2910 MeV \cite{Capstick:1986bm}, which will enhance the above
assignment.

It should be noted that there are no $\Lambda_c$ excitation states
around 2940 MeV in the relativistic quark model
predictions~\cite{Ebert:2007nw}. There, $\Lambda_c(2940)$ was
assigned to be the first radial excited state with $J^P=3/2^+$, of
which the predicted mass was slightly below the experimental
value. As the decay of the radial excited state into the $D^0p $
channel is forbidden in the non-relativistic limit, more elaborate
estimate of the relativistic corrections should be necessary.

\subsection{$\Lambda_c(2765)$}

Experimental information about the $\Lambda_c(2765)$ is much
poorer than $\Lambda_c(2880)$ and $\Lambda_c(2940)$. Thus, we
leave it to be discussed as the last $\Lambda_c$ excitation state.

$\Lambda_c(2765)$ was first observed in $\Lambda_c \pi\pi$ by CLEO
Collaboration \cite{Artuso:2000xy,PDG} with a rather broad width
of about 50 MeV, and appeared to resonate through $\Sigma_c\pi$
and probably also $\Sigma_c (2520)\pi$.  At Belle, its broad
structure stands out clearly in the $\Lambda_c \pi\pi$ invariant
mass spectrum~\cite{Abe:2006rz}. However, almost nothing  about
its quantum numbers is known, including whether it is a
$\Lambda_c$ or a $\Sigma_c$ excitation. Cheng {\it et al.} suggest
that $\Lambda_c(2765)$ could be the first excited state of
$\Lambda_c$ with positive-parity according to the predictions of
Skyrme model \cite{Oh:1995ey} and the quark model
\cite{Capstick:1986bm}. It was also proposed that the
$\Lambda_c(2765)$ could be either the first radial ($1S$)
excitation of the $\Lambda_c$ ($J^P=1/2^+$) with a light scalar
diquark component, or the first orbital excitation ($1P$) of the
$\Sigma_c$ ($J^P=3/2^-$) with a light axial vector diquark
\cite{Ebert:2007nw}.

Interestingly, our calculation shows that $\Lambda_c(2765)$ is
very likely to be a $P_\rho$-mode excitation of $\Lambda_c$. The
reason is that, the masses of the two $^2P_\lambda$-mode
excitations, $\Lambda_c(2593)$ and $\Lambda_c(2625)$, are about
$2600$ MeV, and according to the quark model the energies of
$P_\rho$-mode are $\sim 140$ MeV higher than those of
$P_\lambda$-mode \cite{Copley:1979wj}. An implication from this is
that the mass of $P_\rho$-mode excitation is around $2740$ MeV,
which seems to fit into the mass spectrum well. We calculate the
widths of all possible configurations, and the results are listed
in Tab. \ref{w2765}. Comparing with the experiment data, we find
that $\Lambda_c(2765)$ as a $|\Lambda_c \ ^2
P_\rho\frac{1}{2}^-\rangle$ or $|\Lambda_c \ ^2
P_\rho\frac{3}{2}^-\rangle$ state is excluded due to the much
broader widths. The first radial excitation of the $\Lambda_c$
with $J^P=1/2^+$ is also excluded for its extremely narrow width.

Note that the $\Lambda_c(2765)$ was observed in a similar decay
channel as $\Lambda_c(2880)$, i.e. in $\Lambda_c \pi\pi$, and via
$\Sigma_c\pi/\Sigma_c(2520)\pi$. We hence assume that the decay
modes of $\Lambda_c(2765)$ have a similar behavior as those of
$\Lambda_c(2880)$, except that the $D^0p$ channel is forbidden
since the $\Lambda_c(2765)$ is below the $D^0p$ threshold. One
thus expects that $[\Gamma(\Sigma_c
\pi)+\Gamma(\Sigma_c(2520)\pi)]/\Gamma(\Lambda_c\pi\pi) \sim 0.4$
\cite{PDG}, which is similar to the experimental value of
$\Lambda_c(2880)$. We can then calculate the partial decay width
of $\Lambda_c(2765)\to \Sigma_c \pi$ and $\Sigma_c(2520)\pi$ for
different configurations, and predict its total width into
$\Lambda_c\pi\pi$. For $\Lambda_c(2765)$ being a $|\Lambda_c \ ^4
P_\rho\frac{1}{2}^-\rangle$ state, we obtain $[\Gamma(\Sigma_c
\pi)+\Gamma(\Sigma_c(2520)\pi)]\simeq 21.6$ MeV. Thus,
$\Gamma^{exp}_{total}\simeq 21.6/0.4= 53.4$ MeV is obtained and
agrees well with the experimental value $\Gamma^{exp}\simeq 50\sim
73$ MeV~\cite{Artuso:2000xy,Abe:2006rz}.

For $|\Lambda_c \ ^4 P_\rho\frac{3}{2}^-\rangle$, it shows that
the partial decay width for $\Lambda_c(2765)\to \Sigma_c(2520)\pi$
is much larger than for $\Sigma_c \pi$ by about a factor of 50. If
this is the case, one would expect that $\Sigma_c(2520)\pi$ be the
dominant decay channel which however is not consistent with the
data. For $|\Lambda_c \ ^4 P_\rho\frac{5}{2}^-\rangle$, the
extracted decay widths are rather small to compare with its total
width. The above results make the $\Lambda_c(2765)$ a good
candidate for $|\Lambda_c \ ^4 P_\rho\frac{1}{2}^-\rangle$ state,
which also agrees with the quark model prediction.

We also check the possibility of the $\Lambda_c(2765)$ being a
$\Sigma_c$-type state. As the masses of the $D$-wave
$\Sigma_c$-type states in the $N=2$ shell are generally larger
than 2.8 GeV in the quark
model~\cite{Copley:1979wj,Capstick:1986bm}, and the decay channel
$\Lambda_c \pi$ of $P$-wave states in the $N=2$ shell is forbidden
due to the quark model selection rules (see Tab. \ref{css}), only
the $P$-wave states in the $N=1$ shell and radial excitations are
possible.

We calculate the decay widths for those possible states, and the
results are listed in Tab. \ref{wss2800}. It shows that the radial
excitation should be excluded since the decay width is extremely
narrow. The negative parity states,  except $|\Sigma^{+}_c \ ^2
P_\lambda\frac{3}{2}^-\rangle $,  can produce widths at the same
order of magnitude as the data when sum all the decay channel
together. However, note that the dominant channel of
$\Sigma_c$-type charmed states is $\Lambda_c\pi$. The assignment
of $\Lambda_c(2765)$ to a $\Sigma_c$  excitation  will lead to
apparent contradictions to the experimental observations, thus can
be ruled out.

\subsection{$\Sigma_c(2800)$}

The observation of $\Sigma^{++,+,0}_c(2800)$ by Belle in
$\Lambda_c \pi$ channel enriches the spectrum of $\Sigma_c$
excitation states \cite{Mizuk:2004yu}. However, the present
experimental information still cannot determine its quantum
numbers. Theoretical studies appear strongly model-dependent where
its spin-parity of $J^P=1/2^-$, $3/2^-$, or $5/2^-$, seems
possible~\cite{Cheng:2006dk,Chen:2007xf,Gerasyuta:2007un,Garcilazo:2007eh,Ebert:2007nw}.

Almost all the recent theoretical predictions suggest that
$\Sigma_c(2800)$ could be the first orbital excitations, however,
its quantum numbers are different in different models. Its
spin-parity could be $J^P=3/2^-$ in the heavy hadron chiral
perturbation theory  predictions \cite{Cheng:2006dk}, $J^P=3/2^-$
or $J^P=5/2^-$ in the $^3P_0$ model \cite{Chen:2007xf},
$J^P=5/2^-$ in the relativistic quark
model~\cite{Gerasyuta:2007un}, $J^P=1/2^-$ or $3/2^-$ in the
Faddeev studies \cite{Garcilazo:2007eh}, and $J^P=1/2^-$, $3/2^-$
or $5/2^-$ the latest calculations with the relativistic quark
model \cite{Ebert:2007nw}.

Again, taking the quark model guidance that the masses of the
$D$-wave $\Sigma_c$ excitations in the $N=2$ shell are much larger
than 2800 MeV \cite{Copley:1979wj,Capstick:1986bm}, while the
decay channel $\Lambda_c \pi$ of $P$-wave states in $N=2$ shell is
forbidden (see Tab. \ref{css}), we classify the $\Sigma_c(2880)$
as a $P$-wave state in either the $N=1$ shell (i.e., the first
orbital excitation) or the radial excitation. The decay widths of
$\Lambda_c \pi$, $\Sigma_c \pi$ and $\Sigma_c (2520)\pi$ are
calculated, and the results are listed in Tab. \ref{ws2800}.

The radial excitations can be excluded easily due to the extremely
small predictions of the widths compared with the experimental
data. Furthermore, the $\Lambda_c \pi$ channel may dominate over
other channels since $\Sigma_c(2800)$ was only seen there. Thus,
$|\Sigma^{++}_c \ ^2 P_\lambda\frac{3}{2}^-\rangle $, $|\Sigma_c \
^4 P_\lambda\frac{3}{2}^-\rangle $ and $|\Sigma^{++}_c \ ^4
P_\lambda\frac{1}{2}^-\rangle $ should be ruled out due to the
dominance of  either $\Sigma_c (2520)\pi$ or $\Sigma_c \pi$. After
this, it leaves two possible states, $|\Sigma^{++}_c \ ^2
P_\lambda\frac{1}{2}^-\rangle $ and $|\Sigma^{++}_c \ ^4
P_\lambda\frac{5}{2}^-\rangle $, to be assigned to
$\Sigma_c(2800)$. This comes to the same starting point as other
works~\cite{Cheng:2006dk,Chen:2007xf,Gerasyuta:2007un,Garcilazo:2007eh,Ebert:2007nw},
and indicates how poor we know about this state.

In these two states, $\Sigma_c(2800)$ as a $|\Sigma_c \ ^2
P_\lambda\frac{1}{2}^-\rangle $ state (i.e. a first $P$-wave
orbital $\Sigma_c$ excitation) is favored if there are no other
decay channels to contribute significantly to the total width.
Considering there might exist other decay channels and the
uncertainties of the model, $|\Sigma^{++}_c \ ^4
P_\lambda\frac{5}{2}^-\rangle $ is favored since its decays into
$\Lambda_c\pi$ are the dominant channel, while into $\Sigma_c\pi$
and $\Sigma_c(2520)\pi$ are relatively small. The sum of these
three channels, though smaller than the experimental total width,
is acceptable taking into account the uncertainties. To determine
the quantum number of $\Sigma_c(2800)$, a measurement of the ratio
of $\Lambda_c\pi/\Sigma_c (2520)\pi$ or
$\Sigma_c\pi/\Sigma_c(2520)\pi$, or the $\Lambda_c\pi$ angular
distributions should be useful.

\section{Summary}\label{sum}

In the framework of the non-relativistic quark model, the strong
decays of charmed baryons are analyzed with an effective chiral
Lagrangian for the pseudoscalar-meson-quark coupling. This
framework is successful in reproducing the strong decay widths of
$\Sigma_c\to \Lambda_c \pi$, $\Lambda_c(2593)\to \Sigma_c \pi$ and
$\Lambda_c(2625)\to \Sigma_c \pi$. It allows us to fix an
additional parameter $\delta$ which is introduced to account for
model uncertainties arising from the pseudoscalar-meson-quark
coupling constants. We then carry out calculations for those newly
observed states by assuming their possible configurations in the
quark model. By comparing the theoretical results with the
experimental measurement, we extract information about the
classification of those states and their possible quantum numbers.

To be more specific, our results show that both $\Lambda_c(2880)$
and $\Lambda_c(2940)$ are consistent with being internal $D$-wave
states. For the $\Lambda_c(2880)$, its narrow widths, visible
decays into $D^0p$ and the measured ratio
$\mathcal{R}=\Gamma_c(2520)\pi/\Gamma_c(2455)\pi$ suggest a
favored configuration $|\Lambda_c \ ^2
D_{\lambda\lambda}\frac{3}{2}^+\rangle $ with $l_\lambda=2$ and
$l_\rho=0$. Considering the decay width and decay channel of
$\Lambda_c(2940)$, our results indicate that $\Lambda_c(2940)$
could be a $|\Lambda_c \ ^2 D_{\lambda\lambda}\frac{5}{2}^+\rangle
$ state. Our predictions are different from the suggestions of
Ref. \cite{Cheng:2006dk,Chen:2007xf} that $\Lambda_c(2880)$ is a
$l_\lambda=l_\rho=1$ orbital excition state with $J^P=5/2^+$.
Although the angular distribution fit for
$\Lambda_c(2880)\to\Sigma_c \pi$ favors $J=5/2$ \cite{Abe:2006rz},
the data still possess large uncertainties and more precise
measurements are desired.

We propose that $\Lambda_c(2765)$ is most likely a $\rho$-mode
$P$-wave excitation in the $N=1$ shell. In those multiplets, the
most possible state is $|\Lambda_c \ ^4
P_\rho\frac{1}{2}^-\rangle$, which also turns to be consistent
with the quark model predictions.

For the $\Sigma_c(2800)$, the present experimental information
seems not sufficient for its classification in our approach.
Assuming that no other sizeable decay channels apart from
$\Lambda_c\pi$, $\Sigma_c\pi$ and $\Sigma_c(2520)\pi$, to
contribute to its total width, it is most likely a $|\Sigma_c \ ^2
P_\lambda\frac{1}{2}^-\rangle $ state. Otherwise, the possibility
of its being a $|\Sigma^{++}_c \ ^4 P_\lambda\frac{5}{2}^-\rangle
$ state can not be excluded. Measurements of the ratio of
$\Lambda_c\pi/\Sigma_c(2520)\pi$ and/or
$\Sigma_c\pi/\Sigma_c(2520)\pi$ are recommended to clarify its
spin-parity.

\section*{  Acknowledgement }

This work is supported, in part, by the National Natural Science
Foundation of China (Grants 10675131 and 10775145), Chinese
Academy of Sciences (KJCX3-SYW-N2), the U.K. EPSRC (Grant No.
GR/S99433/01), the Post-Doctoral Programme Foundation of China,
and K. C. Wong Education Foundation, Hong Kong.


\begin{widetext}
\begin{center}
\begin{table}[ht]
\caption{The spatial wave functions with principal quantum number
$N\leq 2$, denoted by $^{N}\Psi^{\sigma}_{ L L_z}$, where
$\sigma=s,\lambda,\rho,A,\rho\rho,\lambda\lambda$ stands for
different excitation modes in quark model. } \label{wffff}
\begin{tabular}{|c|c|c||c|c|c| }\hline\hline
N  &\ \ \textbf{L} L \ \ $l_\rho$\ \ $l_\lambda$ \ \ & \ \ $n_\rho$ $n_\lambda$ \ \ & $L_z$&\ \ wave function\ \   \\
\hline
0& 0 \ \ 0 \ \ 0 \ \ 0    &0 \ \ 0  & 0 & $^{0}\Psi^S_{00}=\psi_{00}(\rho)\psi_{00}(\lambda)$      \\
\hline
1& 1 \ \ 1 \ \ 0 \ \ 1    &0 \ \ 0  & m & $^{1}\Psi^{\lambda}_{1m}=\psi_{00}(\rho)\psi_{1m}(\lambda)$        \\
\hline
1& 1 \ \  1 \ \  1 \ \  0    & 0 \ \ 0 & m& $^{1}\Psi^{\rho}_{1m}=\psi_{1m}(\rho)\psi_{00}(\lambda)$      \\
\hline 2& 2 \ \  0 \ \ 1 \ \ 1    &0 \ \ 0 & 0 &
$^{2}\Psi^{A}_{00}=\sqrt{\frac{1}{3}}R_{0
1}(\rho)R_{01}(\lambda)\{Y_{11}(\lambda)
Y_{1-1}(\rho)-Y_{10}(\lambda)Y_{10}(\rho)+Y_{1-1}(\lambda)Y_{11}(\rho)\}$      \\
\hline 2& 2 \ \  1 \ \ 1 \ \  1    &0 \ \ 0 & 1&
$^{2}\Psi^{A}_{11}=\sqrt{\frac{1}{2}}R_{0
1}(\rho)R_{01}(\lambda)\{Y_{11}(\lambda)
Y_{10}(\rho)-Y_{10}(\lambda)Y_{11}(\rho)\}$        \\
2& 2 \ \ 1 \ \  1 \ \ 1    &0 \ \ 0 & 0&
$^{2}\Psi^{A}_{10}=\sqrt{\frac{1}{2}}R_{0
1}(\rho)R_{01}(\lambda)\{Y_{11}(\lambda)
Y_{1-1}(\rho)-Y_{1-1}(\lambda)Y_{11}(\rho)\}$       \\
2& 2 \ \ 1 \ \ 1 \ \ 1    &0 \ \ 0 & $-1$ &
$^{2}\Psi^{A}_{1-1}=\sqrt{\frac{1}{2}}R_{0
1}(\rho)R_{01}(\lambda)\{Y_{10}(\lambda)
Y_{1-1}(\rho)-Y_{1-1}(\lambda)Y_{10}(\rho)\}$       \\
\hline
\end{tabular}
\end{table}
\end{center}
\end{widetext}

\begin{table}[ht]
\caption{The total wave functions of $\Lambda_c$-type charmed
baryons, denoted by $|\Lambda_c \ ^{2S+1} L_{\sigma}J^P\rangle$ as
used in Ref.\cite{Copley:1979wj}. The Clebsch-Gordan series for
the spin and angular-momentum addition $|\Lambda_c \ ^{2S+1}
L_{\sigma}J^P\rangle= \sum_{L_z+S_z=J_z} \langle LL_z,SS_z|JJ_z
\rangle  ^{N}\Psi^{\sigma }_{LL_z} \chi_{S_z}\phi_{\Lambda_c}$ has
been omitted. } \label{wfL}
\begin{tabular}{|c|c|c|c|c|c|c }\hline\hline
state & N  &\ \ \textbf{J} \ L \ \ $S$\ \ & \ \ $J^P$  \ \ &\ \ wave function   \\
\hline
$|\Lambda_c \ ^2 S\frac{1}{2}^+\rangle$& 0 & \ $\frac{1}{2}$ \ \ 0 \ \ $\frac{1}{2}$ \ \    & $\frac{1}{2}^+$  & $^{0}\Psi^S_{00}\chi^\rho_{S_z}\phi_{\Lambda_c}$      \\
\hline
$|\Lambda_c \ ^2 P_\lambda\frac{1}{2}^-\rangle$&1& \ $\frac{3}{2}$ \ \ 1 \ \ $\frac{1}{2}$ \ \ &$\frac{1}{2}^-$ &  $   ^{1}\Psi^{\lambda}_{1L_z} \chi^{\rho}_{S_z}\phi_{\Lambda_c}$                     \\
$|\Lambda_c \ ^2 P_\lambda\frac{3}{2}^-\rangle$&1& \ $\frac{3}{2}$ \ \ 1 \ \ $\frac{1}{2}$ \ \ &$\frac{3}{2}^-$ &         \\
\hline
$|\Lambda_c \ ^2 P_\rho\frac{1}{2}^-\rangle$&1& \ $\frac{3}{2}$ \ \ 1 \ \ $\frac{1}{2}$ \ \ &$\frac{1}{2}^-$ &  $  ^{1}\Psi^{\rho}_{1L_z} \chi^{\lambda}_{S_z}\phi_{\Lambda_c}$        \\
$|\Lambda_c \ ^2 P_\rho\frac{3}{2}^-\rangle$&1& \ $\frac{3}{2}$ \ \ 1 \ \ $\frac{1}{2}$ \ \ &$\frac{3}{2}^-$ &        \\
\hline
$|\Lambda_c \ ^4 P_\rho\frac{1}{2}^-\rangle$&1& \ $\frac{5}{2}$ \ \ 1 \ \ $\frac{3}{2}$ \ \ &$\frac{1}{2}^-$ &        \\
$|\Lambda_c \ ^4 P_\rho\frac{3}{2}^-\rangle$&1& \ $\frac{5}{2}$ \ \ 1 \ \ $\frac{3}{2}$ \ \ &$\frac{3}{2}^-$ &  $   ^{1}\Psi^{\rho}_{1L_z} \chi^{S}_{S_z}\phi_{\Lambda_c}$      \\
\hline
\end{tabular}
\end{table}

\begin{table}[ht]
\caption{The total wave functions of $\Sigma_c$-type charmed
baryons, denoted by $|\Sigma_c \ ^{2S+1} L_{\sigma}J^P\rangle$.
The Clebsch-Gordan series for the spin and angular-momentum
addition $|\Sigma_c \ ^{2S+1} L_{\sigma}J^P\rangle=
\sum_{L_z+S_z=J_z} \langle LL_z,SS_z|JJ_z \rangle ^{N}\Psi^{\sigma
}_{LL_z} \chi_{S_z}\phi_{\Sigma_c}$ has been omitted.} \label{wfS}
\begin{tabular}{|c|c|c|c|c|c|c }\hline\hline
state & N  &\ \ \textbf{J} \ L \ \ $S$\ \ & \ \ $J^P$  \ \ &\ \ wave function   \\
\hline
$|\Sigma_c \ ^2 S\frac{1}{2}^+\rangle$& 0 & \ $\frac{1}{2}$ \ \ 0 \ \ $\frac{1}{2}$ \ \    & $\frac{1}{2}^+$  & $^{0}\Psi^S_{00}\chi^\lambda_{S_z}\phi_{\Sigma_c}$      \\
\hline
$|\Sigma_c \ ^4 S\frac{3}{2}^+\rangle$& 0 & \ $\frac{3}{2}$ \ \ 0 \ \ $\frac{3}{2}$ \ \    & $\frac{3}{2}^+$  & $^{0}\Psi^S_{00}\chi^s_{S_z}\phi_{\Sigma_c}$      \\
\hline
$|\Sigma_c \ ^2 P_\lambda\frac{1}{2}^-\rangle$&1& \ $\frac{3}{2}$ \ \ 1 \ \ $\frac{1}{2}$ \ \ &$\frac{1}{2}^-$ &  $  ^{1}\Psi^{\lambda}_{1L_z} \chi^{\lambda}_{S_z}\phi_{\Sigma_c}$                     \\
$|\Sigma_c \ ^2 P_\lambda\frac{3}{2}^-\rangle$&1& \ $\frac{3}{2}$ \ \ 1 \ \ $\frac{1}{2}$ \ \ &$\frac{3}{2}^-$ &         \\
\hline
$|\Sigma_c \ ^4 P_\lambda\frac{1}{2}^-\rangle$&1& \ $\frac{5}{2}$ \ \ 1 \ \ $\frac{3}{2}$ \ \ &$\frac{1}{2}^-$ &                       \\
$|\Sigma_c \ ^4 P_\lambda\frac{3}{2}^-\rangle$&1& \ $\frac{5}{2}$ \ \ 1 \ \ $\frac{3}{2}$ \ \ &$\frac{3}{2}^-$ &  $   ^{1}\Psi^{\lambda}_{1L_z} \chi^{s}_{S_z}\phi_{\Sigma_c}$ \\
$|\Sigma_c \ ^4 P_\lambda\frac{5}{2}^-\rangle$&1& \ $\frac{5}{2}$ \ \ 1 \ \ $\frac{3}{2}$ \ \ &$\frac{5}{2}^-$ &                      \\
\hline
\end{tabular}
\end{table}

\begin{table}[ht]
\caption{The spin-factors used in this work.} \label{gfactor}
\begin{tabular}{|l |l| l|c|c|c|c }\hline
\hline
$g_1=\langle \chi^\lambda_{1/2}|\sigma_{1z}|\chi^\rho_{1/2}\rangle=-\frac{1}{\sqrt{3}}$             \     &   $g^{D}_1=\langle \chi^\rho_{1/2}|\sigma_{3z}|\chi^\lambda_{1/2}\rangle=0$   \\
$g_2=\langle \chi^\lambda_{1/2}|\sigma^+_{1}|\chi^\rho_{-1/2}\rangle=-\frac{1}{\sqrt{3}}$            \    &   $g^{D}_2=\langle \chi^\rho_{1/2}|\sigma^+_{3}|\chi^\lambda_{-1/2}\rangle=0$  \\
$g_7=\langle \chi^\lambda_{1/2}|\sigma_{1z}|\chi^s_{1/2}\rangle=\frac{\sqrt{2}}{3}$                  \     &   $g^{D}_3=\langle \chi^\rho_{1/2}|\sigma_{3z}|\chi^s_{1/2}\rangle=0$          \\
$g_{3}=\langle \chi^\lambda_{1/2}|\sigma^-_{1}|\chi^s_{3/2}\rangle=-\frac{1}{\sqrt{6}}$              \     &   $g^{D}_{4}=\langle \chi^\rho_{1/2}|\sigma^+_{3}|\chi^s_{-1/2}\rangle=0$      \\
$g_4=\langle \chi^\lambda_{1/2}|\sigma^+_{1}|\chi^s_{-1/2}\rangle=\frac{\sqrt{2}}{6}$               \    &   $g^{D}_5=\langle \chi^\rho_{1/2}|\sigma^-_{3}|\chi^s_{3/2}\rangle=0$        \\
$g_5=\langle \chi^\lambda_{1/2}|\sigma_{1z}|\chi^\lambda_{1/2}\rangle=\frac{2}{3}$                  \     &  $g^{D}_6=\langle \chi^\rho_{1/2}|\sigma_{3z}|\chi^\rho_{1/2}\rangle=1$            \\
$g_6=\langle \chi^\lambda_{1/2}|\sigma^+_{1}|\chi^\lambda_{-1/2}\rangle=-\frac{2}{3}$               \     &  $g^{D}_7=\langle \chi^\rho_{1/2}|\sigma^+_{3}|\chi^\rho_{-1/2}\rangle=1$          \\
$g^{\Sigma}_1=\langle \chi^\rho_{1/2}|\sigma_{1z}|\chi^\lambda_{1/2}\rangle=-\frac{1}{\sqrt{3}}$    \     &  $g^{*}_1=\langle \chi^s_{3/2}|\sigma^+_{1}|\chi^\lambda_{1/2}\rangle=-\frac{1}{\sqrt{6}}$    \\
\hline
\end{tabular}
\end{table}

\begin{widetext}
\begin{center}
\begin{table}[ht]
\caption{ The decay amplitudes for all the states $|\Lambda_c \
^{2S+1} L_{\sigma}J^P\rangle$ up to $N=2$ shell in $\Sigma_c \pi$
channel (a factor 2 is omitted). $g_I$ is an isospin factor which
is defined by $g_I=\langle \phi_\Sigma|I_1|\phi_\Lambda\rangle$.
$F$, as the decay form factor, is defined in Eq. (\ref{formf}). }
\label{asa}
\begin{tabular}{|c|c|c|c|c|c|c }\hline\hline
initial state                                                                             & amplitude\\
\hline
$|\Lambda_c \ ^2 S\frac{1}{2}^+\rangle $                      &forbidden by the kinematics\\
\hline
$|\Lambda_c \ ^2 P_\lambda\frac{1}{2}^-\rangle      $              &$i\frac{\sqrt{6}}{6}g_1g_I\left\{Gq-\left(\frac{\sqrt{6}}{12}q_\lambda+\frac{\sqrt{2}}{4}q_\rho\right)h\right\}\frac{q_\lambda}{\alpha_\lambda}F+i\frac{1}{2}g_1g_Ih\alpha_\lambda F$\\
$|\Lambda_c \ ^2 P_\lambda\frac{3}{2}^-\rangle      $              &$-i\frac{\sqrt{3}}{3}g_1g_I\left\{Gq-\left(\frac{\sqrt{6}}{12}q_\lambda+\frac{\sqrt{2}}{4}q_\rho\right)h\right\}\frac{q_\lambda}{\alpha_\lambda}F$ \\
\hline
$|\Lambda_c \ ^2 P_\rho\frac{1}{2}^-\rangle         $                 &$i\frac{\sqrt{6}}{6}g_5g_I\left\{Gq-\left(\frac{\sqrt{6}}{12}q_\lambda+\frac{\sqrt{2}}{4}q_\rho\right)h\right\}\frac{q_\rho}{\alpha_\rho}F+i\frac{\sqrt{3}}{6}(g_5+2g_6)g_Ih\alpha_\rho F$\\
$|\Lambda_c \ ^2 P_\rho\frac{3}{2}^-\rangle         $                 &$-i\frac{\sqrt{3}}{3}g_5g_I\left\{Gq-\left(\frac{\sqrt{6}}{12}q_\lambda+\frac{\sqrt{2}}{4}q_\rho\right)h\right\}\frac{q_\rho}{\alpha_\rho}F-i\frac{\sqrt{6}}{6}(g_5-g_6)g_Ih\alpha_\rho F$\\
\hline
$|\Lambda_c \ ^4 P_\rho\frac{1}{2}^-\rangle         $                 &$-i\frac{\sqrt{6}}{6}g_7g_I\left\{Gq-\left(\frac{\sqrt{6}}{12}q_\lambda+\frac{\sqrt{2}}{4}q_\rho\right)h\right\}\frac{q_\rho}{\alpha_\rho}F-i\frac{\sqrt{3}}{6}\left(\sqrt{3}g_3-g_4-g_7\right)g_Ih\alpha_\rho F$ \\
$|\Lambda_c \ ^4 P_\rho\frac{3}{2}^-\rangle         $                 &$-i\frac{\sqrt{30}}{30}g_7g_I\left\{Gq-\left(\frac{\sqrt{6}}{12}q_\lambda+\frac{\sqrt{2}}{4}q_\rho\right)h\right\}\frac{q_\rho}{\alpha_\rho}F-i\frac{\sqrt{15}}{30}\left(g_7+2\sqrt{3}g_3+4g_4\right)g_Ih\alpha_\rho F$ \\
\hline
\end{tabular}
\end{table}
\end{center}
\end{widetext}

\begin{widetext}
\begin{center}
\begin{table}[ht]
\caption{ The decay amplitudes for all the states $|\Lambda_c \
^{2S+1} L_{\sigma}J^P\rangle$ up to $N=2$ shell in $\Sigma_c(2520)
\pi$ channel (a factor 2 is omitted). $F$, as the decay form
factor, is defined in Eq. (\ref{formf}).} \label{asb}
\begin{tabular}{|c|c|c|c|c|c|c }\hline\hline
initial state                                                                     &  $J_z$       & amplitude\\
\hline
$|\Lambda_c \ ^2 S_A \frac{1}{2}^+\rangle$             & $\frac{1}{2}$        &$\frac{\sqrt{3}}{6}g^{*}_2g_I\left\{Gq-\left(\frac{\sqrt{6}}{12}q_\lambda+\frac{\sqrt{2}}{4}q_\rho\right)h\right\}\frac{q_\lambda}{\alpha_\lambda}\frac{q_\rho}{\alpha_\rho}F+\frac{\sqrt{2}}{12}g^{*}_2g_Ih\left(\alpha_\lambda\frac{q_\rho}{\alpha_\rho}+\sqrt{3}\alpha_\rho\frac{q_\lambda}{\alpha_\lambda} \right)F$\\
\hline
$|\Lambda_c \ ^4 S_A \frac{3}{2}^+\rangle$             & $\frac{1}{2}$    &$\frac{\sqrt{3}}{6}g^{*}_7g_I\left\{Gq-\left(\frac{\sqrt{6}}{12}q_\lambda+\frac{\sqrt{2}}{4}q_\rho\right)h\right\}\frac{q_\lambda}{\alpha_\lambda}\frac{q_\rho}{\alpha_\rho}F+\frac{\sqrt{2}}{12}g^{*}_7g_Ih\left(\alpha_\lambda\frac{q_\rho}{\alpha_\rho}+\sqrt{3}\alpha_\rho\frac{q_\lambda}{\alpha_\lambda} \right)F$\\
                                                       & $\frac{3}{2}$    &$\frac{\sqrt{3}}{6}g^{*}_4g_I\left\{Gq-\left(\frac{\sqrt{6}}{12}q_\lambda+\frac{\sqrt{2}}{4}q_\rho\right)h\right\}\frac{q_\lambda}{\alpha_\lambda}\frac{q_\rho}{\alpha_\rho}F+\frac{\sqrt{2}}{12}g^{*}_4g_Ih\left(\alpha_\lambda\frac{q_\rho}{\alpha_\rho}+\sqrt{3}\alpha_\rho\frac{q_\lambda}{\alpha_\lambda} \right)F$\\
\hline
$|\Lambda_c \ ^2 P_A \frac{1}{2}^-\rangle$             & $\frac{1}{2}$       & $\frac{1}{6}g^{*}_3g_Ih\left(\alpha_\lambda\frac{q_\rho}{\alpha_\rho}-\sqrt{3}\alpha_\rho\frac{q_\lambda}{\alpha_\lambda} \right)F$\\
\hline
$|\Lambda_c \ ^2 P_A \frac{3}{2}^-\rangle$             & $\frac{1}{2}$     & $\frac{\sqrt{2}}{12}g^{*}_3g_Ih\left(\alpha_\lambda\frac{q_\rho}{\alpha_\rho}-\sqrt{3}\alpha_\rho\frac{q_\lambda}{\alpha_\lambda} \right)F$\\
                                                       & $\frac{3}{2}$    & $\frac{\sqrt{6}}{12}g^{*}_1g_Ih\left(\alpha_\lambda\frac{q_\rho}{\alpha_\rho}-\sqrt{3}\alpha_\rho\frac{q_\lambda}{\alpha_\lambda} \right)F$\\
\hline
\end{tabular}
\end{table}
\end{center}
\end{widetext}

\begin{widetext}
\begin{center}
\begin{table}[ht]
\caption{The decay amplitudes for all the states $|\Sigma_c \
^{2S+1} L_{\sigma}J^P\rangle$ up to $N=2$ shell in $\Lambda_c \pi$
channel (a factor 2 is omitted). $F$, as the decay form factor, is
defined in Eq. (\ref{formf}).} \label{css}
\begin{tabular}{|c|c|c|c|c|c|c }\hline\hline
initial state                                                                       &   amplitude\\
\hline
$|\Sigma_c \ ^2 S\frac{1}{2}^+\rangle                      $ &$g^{\Sigma}_1g_I\left\{Gq-\left(\frac{\sqrt{6}}{12}q_\lambda+\frac{\sqrt{2}}{4}q_\rho\right)h\right\} F $\\
\hline
$|\Sigma_c \ ^4 S\frac{3}{2}^+\rangle                     $ &$g^{\Sigma}_3g_I\left\{Gq-\left(\frac{\sqrt{6}}{12}q_\lambda+\frac{\sqrt{2}}{4}q_\rho\right)h\right\} F $\\
\hline
$|\Sigma_c \ ^2 P_\lambda\frac{1}{2}^-\rangle             $ &$i\frac{\sqrt{6}}{6}g^{\Sigma}_1g_I\left\{Gq-\left(\frac{\sqrt{6}}{12}q_\lambda+\frac{\sqrt{2}}{4}q_\rho\right)h\right\}\frac{q_\lambda}{\alpha_\lambda}F+i\frac{1}{6}(g^{\Sigma}_1+2g^{\Sigma}_2)g_Ih\alpha_\lambda F$\\
$|\Sigma_c \ ^2 P_\lambda\frac{3}{2}^-\rangle             $ &$-i\frac{\sqrt{3}}{3}g^{\Sigma}_1g_I\left\{Gq-\left(\frac{\sqrt{6}}{12}q_\lambda+\frac{\sqrt{2}}{4}q_\rho\right)h\right\}\frac{q_\lambda}{\alpha_\lambda}F-i\frac{\sqrt{2}}{6}(g^{\Sigma}_1-g^{\Sigma}_2)g_Ih\alpha_\lambda F$\\
\hline
$|\Sigma_c \ ^4 P_\lambda\frac{1}{2}^-\rangle             $ &$i\frac{\sqrt{6}}{6}g^{\Sigma}_3g_I\left\{Gq-\left(\frac{\sqrt{6}}{12}q_\lambda+\frac{\sqrt{2}}{4}q_\rho\right)h\right\}\frac{q_\lambda}{\alpha_\lambda}F+i\frac{1}{6}(g^{\Sigma}_3+g^{\Sigma}_4-\sqrt{3}g^{\Sigma}_5)g_Ih\alpha_\lambda F$\\
$|\Sigma_c \ ^4 P_\lambda\frac{3}{2}^-\rangle             $  &$-i\frac{\sqrt{30}}{30}g^{\Sigma}_3g_I\left\{Gq-\left(\frac{\sqrt{6}}{12}q_\lambda+\frac{\sqrt{2}}{4}q_\rho\right)h\right\}\frac{q_\lambda}{\alpha_\lambda}F-i\frac{\sqrt{5}}{30}(g^{\Sigma}_3+2\sqrt{3}g^{\Sigma}_5+4g^{\Sigma}_4)g_Ih\alpha_\lambda F$\\
$|\Sigma_c \ ^4 P_\lambda\frac{5}{2}^-\rangle            $  &$-i\frac{\sqrt{30}}{10}g^{\Sigma}_3g_I\left\{Gq-\left(\frac{\sqrt{6}}{12}q_\lambda+\frac{\sqrt{2}}{4}q_\rho\right)h\right\}\frac{q_\lambda}{\alpha_\lambda}F-i\frac{\sqrt{15}}{30}(\sqrt{3}g^{\Sigma}_3-\sqrt{3}g^{\Sigma}_4+g^{\Sigma}_5)g_Ih\alpha_\lambda F$\\
\hline
\end{tabular}
\end{table}
\end{center}
\end{widetext}

\begin{widetext}
\begin{center}
\begin{table}[ht]
\caption{The decay amplitudes for the first orbital and radial
excitations $|\Sigma_c \ ^{2S+1} L_{\sigma}J^P\rangle$ in $
\Sigma_c \pi$ channel (a factor 2 is omitted). $F$, as the decay
form factor, is defined in Eq. (\ref{formf}).} \label{F12}
\begin{tabular}{|c|c|c|c|c|c|c }\hline\hline
initial state                                                                       &   amplitude\\
\hline
$|\Sigma_c \ ^2 P_\lambda\frac{1}{2}^-\rangle         $     &$i\frac{\sqrt{6}}{6}g_5g_I\left\{Gq-\left(\frac{\sqrt{6}}{12}q_\lambda+\frac{\sqrt{2}}{4}q_\rho\right)h\right\}\frac{q_\lambda}{\alpha_\lambda}F+i\frac{1}{6}(g_5+2g_6)g_Ih\alpha_\lambda F$\\
$|\Sigma_c \ ^2 P_\lambda\frac{3}{2}^-\rangle         $     &$-i\frac{\sqrt{3}}{3}g_5g_I\left\{Gq-\left(\frac{\sqrt{6}}{12}q_\lambda+\frac{\sqrt{2}}{4}q_\rho\right)h\right\}\frac{q_\lambda}{\alpha_\lambda}F-i\frac{\sqrt{2}}{6}(g_5-g_6)g_Ih\alpha_\lambda F$\\
\hline
$|\Sigma_c \ ^4 P_\lambda\frac{1}{2}^-\rangle         $     &$i\frac{\sqrt{6}}{6}g_7g_I\left\{Gq-\left(\frac{\sqrt{6}}{12}q_\lambda+\frac{\sqrt{2}}{4}q_\rho\right)h\right\}\frac{q_\lambda}{\alpha_\lambda}F+i\frac{1}{6}(g_7+g_4-\sqrt{3}g_3)g_Ih\alpha_\lambda F$\\
$|\Sigma_c \ ^4 P_\lambda\frac{3}{2}^-\rangle         $      &$-i\frac{\sqrt{30}}{30}g_7g_I\left\{Gq-\left(\frac{\sqrt{6}}{12}q_\lambda+\frac{\sqrt{2}}{4}q_\rho\right)h\right\}\frac{q_\lambda}{\alpha_\lambda}F-i\frac{\sqrt{5}}{30}(g_7+2\sqrt{3}g_3+4g_4)g_Ih\alpha_\lambda F$\\
$|\Sigma_c \ ^4 P_\lambda\frac{5}{2}^-\rangle         $      &$-i\frac{\sqrt{30}}{10}g_7g_I\left\{Gq-\left(\frac{\sqrt{6}}{12}q_\lambda+\frac{\sqrt{2}}{4}q_\rho\right)h\right\}\frac{q_\lambda}{\alpha_\lambda}F-i\frac{\sqrt{15}}{30}(\sqrt{3}g_7-\sqrt{3}g_4+g_3)g_Ih\alpha_\lambda F$\\
\hline
$|\Sigma_c \ ^2 S_{\rho\rho}\frac{1}{2}^+\rangle      $     &$\frac{\sqrt{6}}{12}g_5g_I\left\{Gq-\left(\frac{\sqrt{6}}{12}q_\lambda+\frac{\sqrt{2}}{4}q_\rho\right)h\right\}\left(\frac{q_\rho}{\alpha_\rho}\right)^2F+\frac{\sqrt{3}}{6}g_5g_Ihq_\rho F$\\
 \hline
$|\Sigma_c \ ^4 S_{\rho\rho}\frac{3}{2}^+\rangle      $     &$\frac{\sqrt{6}}{12}g_7g_I\left\{Gq-\left(\frac{\sqrt{6}}{12}q_\lambda+\frac{\sqrt{2}}{4}q_\rho\right)h\right\}\left(\frac{q_\rho}{\alpha_\rho}\right)^2F+\frac{\sqrt{3}}{6}g_7g_Ihq_\rho F$\\
\hline
\end{tabular}
\end{table}
\end{center}
\end{widetext}

\begin{widetext}
\begin{center}
\begin{table}[ht]
\caption{ The decay amplitudes for the first orbital and radial
excitations $|\Sigma_c \ ^{2S+1} L_{\sigma}J^P\rangle$ in
$\Sigma_c(2520) \pi$ (a factor 2 is omitted). $F$, as the decay
form factor, is defined in Eq. (\ref{formf}).} \label{f11}
\begin{tabular}{|c|c|c|c|c|c|c }\hline\hline
initial state                                                                     &  $J_z$       & amplitude\\
\hline
$|\Sigma_c \ ^2 P_\lambda\frac{1}{2}^-\rangle$             & $\frac{1}{2}$     &$i\frac{\sqrt{6}}{6}g^{*}_2g_I\left\{Gq-\left(\frac{\sqrt{6}}{12}q_\lambda+\frac{\sqrt{2}}{4}q_\rho\right)h\right\}\frac{q_\lambda}{\alpha_\lambda}F+i\frac{1}{6}(g^{*}_2+2g^{*}_3)g_Ih\alpha_\lambda F$\\
\hline
$|\Sigma_c \ ^2 P_\lambda\frac{3}{2}^-\rangle$             & $\frac{1}{2}$     &$-i\frac{\sqrt{3}}{3}g^{*}_2g_I\left\{Gq-\left(\frac{\sqrt{6}}{12}q_\lambda+\frac{\sqrt{2}}{4}q_\rho\right)h\right\}\frac{q_\lambda}{\alpha_\lambda}F-i\frac{\sqrt{2}}{6}(g^{*}_2-g^{*}_3)g_Ih\alpha_\lambda F$\\
                                                           & $\frac{3}{2}$     &$i\frac{1}{\sqrt{6}}g^{*}_1g_Ih\alpha_\lambda F$\\
\hline
$|\Sigma_c \ ^4 P_\lambda\frac{1}{2}^-\rangle$             & $\frac{1}{2}$     & $i\frac{\sqrt{6}}{6}g^{*}_7g_I\left\{Gq-\left(\frac{\sqrt{6}}{12}q_\lambda+\frac{\sqrt{2}}{4}q_\rho\right)h\right\}\frac{q_\lambda}{\alpha_\lambda}F+i\frac{1}{6}(g^{*}_7+g^{*}_8-\sqrt{3}g^{*}_6)g_Ih\alpha_\lambda F$\\
\hline
$|\Sigma_c \ ^4 P_\lambda\frac{3}{2}^-\rangle$             & $\frac{1}{2}$     & $-i\frac{\sqrt{30}}{30}g^{*}_7g_I\left\{Gq-\left(\frac{\sqrt{6}}{12}q_\lambda+\frac{\sqrt{2}}{4}q_\rho\right)h\right\}\frac{q_\lambda}{\alpha_\lambda}F-i\frac{\sqrt{5}}{30}(g^{*}_7+2\sqrt{3}g^{*}_6+4g^{*}_8)g_Ih\alpha_\lambda F$\\
                                                           & $\frac{3}{2}$     & $-i\frac{\sqrt{30}}{10}g^{*}_4g_I\left\{Gq-\left(\frac{\sqrt{6}}{12}q_\lambda+\frac{\sqrt{2}}{4}q_\rho\right)h\right\}\frac{q_\lambda}{\alpha_\lambda}F-i\frac{\sqrt{15}}{30}(\sqrt{3}g^{*}_4+2g^{*}_5)g_Ih\alpha_\lambda F$\\
\hline
\end{tabular}
\end{table}
\end{center}
\end{widetext}

\begin{widetext}
\begin{center}
\begin{table}[ht]
\caption{The amplitudes of $\Lambda_c$-type charmed baryons decay
into $D^0p$. $F(q'_\lambda)=\exp { (q'^{2}_\lambda
/4\alpha^2_\lambda  )} $ is the form factor.} \label{adp}
\begin{tabular}{|c|c|c|c|c|c|c }\hline\hline
initial state                                                                              & amplitude\\
\hline
$|\Lambda_c \ ^2 D_{\lambda\lambda}\frac{3}{2}^+\rangle $     & $\frac{\sqrt{15}}{30}\left(G q-\frac{\sqrt{6}}{6}h q'_\lambda\right)\left(\frac{q'_\lambda}{\alpha_\lambda}\right)^2F(q'_\lambda)+\frac{\sqrt{10}}{6}hq'_\lambda F(q'_\lambda)$  \\
$|\Lambda_c \ ^2 D_{\lambda\lambda}\frac{5}{2}^+\rangle $     & $-\frac{\sqrt{10}}{20}\left(G q-\frac{\sqrt{6}}{6}h q'_\lambda\right)\left(\frac{q'_\lambda}{\alpha_\lambda}\right)^2F(q'_\lambda)$  \\
\hline
$|\Lambda_c \ ^2 S_{\lambda\lambda}\frac{1}{2}^+\rangle $     & $\frac{\sqrt{3}}{12}\left(G q-\frac{\sqrt{6}}{6}h q'_\lambda\right)\left(\frac{q'_\lambda}{\alpha_\lambda}\right)^2F(q'_\lambda)+\frac{\sqrt{2}}{6}hq'_\lambda F(q'_\lambda)$ \\
\hline
\end{tabular}
\end{table}
\end{center}
\end{widetext}

\begin{table}[ht]
\caption{The decay widths for the low-lying charmed baryons.
$\Lambda_c(2593)$ and $\Lambda_c(2625)$ assigned as both
$P_\lambda$- and $P_\rho$-mode excitations are listed. The partial
decay widths for $\Sigma_c $ and $\Sigma_c(2520)\to \Lambda_c\pi$
are also listed. They serve as experimental input for the
determination of parameter $\delta$ in this approach. }
\label{wssss}
\begin{tabular}{|c|c|c|c|c|}
\hline\hline
              & notation  & channel   & $\Gamma_{exp}$ (MeV) & $\Gamma_{th}$ (MeV)  \\
\hline \hline
     $\Lambda_c(2593)$    &$|\Lambda_c \ ^2 P_\lambda\frac{1}{2}^-\rangle$   &  $\Sigma^{++}_c\pi^-$       &  $0.65^{+0.41}_{-0.31}$ &0.37                      \\
                          &                                                  &  $\Sigma^{+}_c\pi^0$        &                         &0.73                      \\
                          &                                                  &  $\Sigma^{0}_c\pi^+$        &  $0.67^{+0.41}_{-0.31}$ &0.40                      \\
\hline
     $\Lambda_c(2625)$    &$|\Lambda_c \ ^2 P_\lambda\frac{3}{2}^-\rangle$   &  $\Sigma^{++}_c\pi^-$       & $<0.10$                 &$1.47\times 10^{-2}$    \\
                          &                                                  &  $\Sigma^{+}_c\pi^0$        &                         &$2.08\times 10^{-2}$    \\
                          &                                                  &  $\Sigma^{0}_c\pi^+$        & $<0.09$                 &$1.50\times 10^{-2}$    \\
\hline
     $\Lambda_c(2593)$   &$|\Lambda_c \ ^2 P_\rho\frac{1}{2}^-\rangle$       &  $\Sigma^{++}_c\pi^-$       & $0.65^{+0.41}_{-0.31}$  &1.02                    \\
\hline
\end{tabular}
\end{table}

\begin{table}[ht]
\caption{The decay widths of $\Lambda_c(2880)$ for all the
possible states in $N=2$  shell (in MeV). Ratio $\mathcal{R}$ is
defined as
$\mathcal{R}=\Gamma(\Sigma_c(2520)\pi^{\pm})/\Gamma(\Sigma_c\pi^{\pm})$}
\label{w2880}
\begin{tabular}{|c|c|c|c|c|c|c }\hline\hline
assignment &$\Sigma^+_c\pi^0$&$\Sigma^{0,++}_c\pi^{+,-}$&$\Sigma_c(2520)\pi^{\pm}$& $\mathcal{R}$ & $D^0p$\\
\hline
$|\Lambda_c \ ^2 S_A \frac{1}{2}^+\rangle$                  &0.45  &  0.47      & 0.29      & 0.62   &0\\
\hline
$|\Lambda_c \ ^4 S_A \frac{3}{2}^+\rangle$                  &0.11  &  0.12      & 0.40      & 3.33   & 0\\
\hline
$|\Lambda_c \ ^2 P_A \frac{1}{2}^-\rangle$                  &0.41  &  0.40      & 0.03      & 0.08   & 0\\
$|\Lambda_c \ ^2 P_A \frac{3}{2}^-\rangle$                  &0.10  &  0.10      & 0.06      & 0.60    & 0\\
\hline
$|\Lambda_c \ ^4 P_A\frac{1}{2}^-\rangle  $                 &0.21  &  0.20      & 0.01      & 0.05   & 0\\
$|\Lambda_c \ ^4 P_A\frac{3}{2}^-\rangle  $                 &0.13  &  0.13      & 0.05      & 0.38   &0 \\
$|\Lambda_c \ ^4 P_A\frac{5}{2}^-\rangle  $                 & 0    &   0        & 0.12      &        &0\\
\hline
\end{tabular}
\end{table}

\begin{table}[ht]
\caption{The decay widths of $\Lambda_c(2940)$ for all the
possible states in $N=2$  shell (in MeV). Ratio $\mathcal{R}$ is
defined as
$\mathcal{R}=\Gamma(\Sigma_c(2520)\pi^{\pm})/\Gamma(\Sigma_c\pi^{\pm})$.}
\label{w2940}
\begin{tabular}{|c|c|c|c|c|c|c }\hline\hline
assignment &$\Sigma^+_c\pi^0$&$\Sigma^{0,++}_c\pi^{+,-}$&$\Sigma_c(2520)\pi^{\pm}$&$\mathcal{R}$& $D^0p$\\
\hline
$|\Lambda_c \ ^2 S_A \frac{1}{2}^+\rangle$                  &0.16    & 0.18       & 0.25      & 1.39     &0\\
\hline
$|\Lambda_c \ ^4 S_A \frac{3}{2}^+\rangle$                  &0.04    & 0.05       & 0.35      & 7.00     & 0\\
\hline
$|\Lambda_c \ ^2 P_A \frac{1}{2}^-\rangle$                  &0.65    & 0.64       & 0.47      & 0.73     & 0\\
$|\Lambda_c \ ^2 P_A \frac{3}{2}^-\rangle$                  &0.16    & 0.16       & 0.12      & 0.75     & 0\\
\hline
$|\Lambda_c \ ^4 P_A\frac{1}{2}^-\rangle  $                 &0.32    & 0.32       & 0.02      & 0.06     & 0\\
$|\Lambda_c \ ^4 P_A\frac{3}{2}^-\rangle  $                 &0.20    & 0.20       & 0.09      & 0.45     &0 \\
$|\Lambda_c \ ^4 P_A\frac{5}{2}^-\rangle  $                 & 0      &   0        & 0.21      &          &0 \\
\hline
\end{tabular}
\end{table}

\begin{table}[ht]
\caption{The decay widths (in MeV) of $\Lambda_c(2765)$ for the
possible excitation modes.
$\Gamma_{\mathrm{sum}}=\Gamma_{\Sigma_c\pi}+\Gamma_{\Sigma^*_c\pi}$,
where $\Sigma^*_c$ stands for $\Sigma_c(2520)$.  } \label{w2765}
\begin{tabular}{|c|c|c|c|c|c|c }\hline\hline
                           assignment                                         &  $\Sigma_c\pi^{+,-,0}$  & $\Sigma^*_c\pi^{+,-,0}$  &  $\Gamma_{\mathrm{sum}}$      & $\Gamma_{total}^{exp}$      \\
\hline \hline
                          $|\Lambda_c \ ^4 P_\rho\frac{1}{2}^-\rangle$      &  7.0                    &    0.2                        &  21.6      &  $\sim50-73$                    \\
                          $|\Lambda_c \ ^4 P_\rho\frac{3}{2}^-\rangle$      &  0.4                    &    20.6                       &  63        &                         \\
                          $|\Lambda_c \ ^4 P_\rho\frac{5}{2}^-\rangle$      &  2.2                    &    0.6                        &  8.4       &                        \\
\hline
                          $|\Lambda_c \ ^2 P_\rho\frac{1}{2}^-\rangle$      &  41.4                   &    3.0                        &  133.2     &                        \\
                          $|\Lambda_c \ ^2 P_\rho\frac{3}{2}^-\rangle$      &  29.8                   &    11.4                       &  123.6     &                        \\
\hline
                         $|\Lambda_c \ ^2 S_{\lambda\lambda}\frac{1}{2}^+\rangle $      &  0.08                 &    0.1                      &  0.5     &                        \\
                         $|\Lambda_c \ ^2 S_{\rho\rho}\frac{1}{2}^+\rangle $      &  0.3                   &    0.3                       &  1.8     &                        \\
                          \hline \hline
\end{tabular}
\end{table}

\begin{table}[ht]
\caption{The decay widths (in MeV) of $\Lambda_c(2765)$ as a
$\Sigma_c$-type excitation.
$\Gamma_{\mathrm{sum}}=\Gamma_{\Lambda_c
\pi}+\Gamma_{\Sigma_c\pi}+\Gamma_{\Sigma^*_c\pi}$, where
$\Sigma^*_c$ stands for $\Sigma_c(2520)$.} \label{wss2800}
\begin{tabular}{|c|c|c|c|c|c|c |c}\hline\hline
 assignment         &  $\Lambda_c \pi^{0}$ &  $\Sigma^{++,0}_c \pi^{-,+}$     &  $ \Sigma^{*++,0}_c \pi^{-,+}$  & $\Gamma_{\mathrm{sum}}$   &$\Gamma^{exp} _{total}$\\
\hline
$|\Sigma^{+}_c \ ^2 P_\lambda\frac{3}{2}^-\rangle $             &     0.02            &  1.98          &  1.40           & 6.78   &$\sim50- 73$    \\
$|\Sigma^{+}_c \ ^2 P_\lambda\frac{1}{2}^-\rangle $             &     36.90           &  9.66          &   1.06          & 58.34   & \\
\hline
$|\Sigma^{+}_c \ ^4 P_\lambda\frac{1}{2}^-\rangle $             &     6.68            &   4.20         &   0.03          & 15.14   &\\
$|\Sigma^{+}_c \ ^4 P_\lambda\frac{3}{2}^-\rangle $             &     3.36            &   0.13         &   8.67          & 20.96   & \\
$|\Sigma^{+}_c \ ^4 P_\lambda\frac{5}{2}^-\rangle $             &     20.16           &   0.75         &   0.60          & 22.86   &\\
\hline
$|\Sigma^{+}_c \ ^2 S_{\rho\rho}\frac{1}{2}^+ \rangle  $        &     0.37            &   0.41         &   0.08          & 1.35    &\\
\hline \hline
\end{tabular}
\end{table}

\begin{table}[ht]
\caption{The decay widths (in MeV) of $\Sigma_c (2800)$ for the
possible excitation modes.
$\Gamma_{\mathrm{sum}}=\Gamma_{\Lambda_c
\pi}+\Gamma_{\Sigma_c\pi}+\Gamma_{\Sigma^*_c\pi}$, where
$\Sigma^*_c$ stands for $\Sigma_c(2520)$.} \label{ws2800}
\begin{tabular}{|c|c|c|c|c|c|c |c}\hline\hline
  assignment         &  $\Lambda_c \pi^{+}$ &  $\Sigma^{++,+}_c \pi^{0,+}$     &  $ \Sigma^{*++,+}_c \pi^{0,+}$  & $\Gamma_{\mathrm{sum}}$   &$\Gamma^{exp} _{total}$\\
\hline
$|\Sigma^{++}_c \ ^2 P_\lambda\frac{3}{2}^-\rangle $           &     0.36            &  1.56          &  2.07           & 7.62  &$75^{+22}_{-17}$    \\
$|\Sigma^{++}_c \ ^2 P_\lambda\frac{1}{2}^-\rangle $           &     46.59           &  14.66         &   1.05          & 78.01  & \\
\hline
$|\Sigma^{++}_c \ ^4 P_\lambda\frac{1}{2}^-\rangle $           &     4.36            &   4.23         &   0.34          & 13.5    &\\
$|\Sigma^{++}_c \ ^4 P_\lambda\frac{3}{2}^-\rangle $           &     4.51            &   0.22         &   10.79         & 26.53  & \\
$|\Sigma^{++}_c \ ^4 P_\lambda\frac{5}{2}^-\rangle $           &     27.08           &   1.34         &   1.41          & 32.58   &\\
\hline
$|\Sigma^{++}_c \ ^2 S_{\rho\rho}\frac{1}{2}^+ \rangle  $      &     0.26            &   0.52         &   0.13          & 1.56    &\\
\hline \hline
\end{tabular}
\end{table}

\end{document}